\documentclass[10pt,letterpaper]{article}

\usepackage{amsmath,amssymb}

\usepackage{changepage}

\usepackage[utf8x]{inputenc}

\usepackage{textcomp,marvosym}

\usepackage{cite}

\usepackage{nameref,hyperref}

\usepackage[right]{lineno}

\usepackage{microtype}
\DisableLigatures[f]{encoding = *, family = * }

\usepackage[table]{xcolor}

\usepackage{array}

\newcolumntype{+}{!{\vrule width 2pt}}

\newlength\savedwidth

\usepackage[aboveskip=1pt,labelfont=bf,labelsep=period,justification=raggedright,singlelinecheck=off]{caption}

\makeatletter
\renewcommand{\@biblabel}[1]{\quad#1.}
\makeatother

\date{}

\usepackage{lastpage,fancyhdr,graphicx}
\usepackage{fullpage}
\usepackage{epstopdf}

\usepackage{amsfonts,amsthm,MnSymbol}

\usepackage{multirow}

\usepackage{color}

\newcommand{\uvec}{\mathbf{u}}

\newcommand{\xvec}{\mathbf{x}}

\newcommand{\zvec}{\mathbf{z}}

\newcommand{\Cvec}{\mathbf{C}}

\newcommand{\Ivec}{\mathbf{I}}
\newcommand{\Jvec}{\mathbf{J}}
\newcommand{\Kvec}{\mathbf{K}}
\newcommand{\Lvec}{\mathbf{L}}

\newcommand{\Pvec}{\mathbf{P}}
\newcommand{\Rvec}{\mathbf{R}}
\newcommand{\Svec}{\mathbf{S}}

\newcommand{\Wvec}{\mathbf{W}}

\newcommand{\Lamvec}{\mathbf{\Lambda}}

\DeclareMathOperator{\Ex}{E}

\DeclareMathOperator{\Var}{Var}
\DeclareMathOperator{\Cov}{Cov}

\begin{document}
\vspace*{0.2in}

\begin{flushleft}
{\Large
\textbf\newline{When do Correlations Increase with Firing Rates?} 
}
\newline
\\
Andrea K. Barreiro \textsuperscript{1*} and
Cheng Ly \textsuperscript{2}
\\
\bigskip
\textbf{1} Department of Mathematics, Southern Methodist University, Dallas, TX 75275  U.S.A.
\\
\textbf{2} Department of Statistical Sciences and Operations Research, Virginia Commonwealth University, Richmond, VA 23284  U.S.A.
\\
\bigskip

%
%





* abarreiro@smu.edu

\end{flushleft}
\section*{Abstract}
A central question in neuroscience is to understand how noisy firing patterns are used to transmit information. 
Because neural spiking is noisy, spiking patterns are often quantified via pairwise correlations, or the probability that two cells will spike coincidentally, 
above and beyond their baseline firing rate. One observation frequently made in experiments, is that correlations can increase systematically with firing rate.  
Theoretical studies have determined that stimulus-dependent correlations that increase with firing rate can have beneficial effects on information coding; however, we still have an incomplete understanding of what circuit mechanisms do, or do not, produce this correlation-firing rate relationship. 
%
%

Here, we study the relationship between pairwise correlations and firing rates in
 recurrently coupled excitatory-inhibitory spiking networks with conductance-based synapses. 
We found that with stronger excitatory coupling, 
a positive relationship emerges between pairwise correlations and firing rates. 
To explain these findings, we used linear response theory to predict the full correlation matrix and to decompose correlations in terms of graph motifs.
We then used this decomposition to explain why covariation of correlations with firing rate --- a relationship previously explained in  feedforward networks driven by correlated input
 --- emerges in some recurrent networks but not in others. 
Furthermore, when correlations covary with firing rate, this relationship is reflected in low-rank structure in the correlation matrix.

%
%

\section*{Author Summary}
A central question in neuroscience is to understand how noisy firing patterns are used to transmit information. 
We quantify spiking patterns by using pairwise \textit{correlations}, or the probability that two cells will spike coincidentally, above and beyond their baseline firing rate. 
One observation frequently made in experiments is that correlations can increase systematically with firing rate.  
Recent studies of a type of output cell in mouse retina found this relationship; furthermore, they determined that the increase of correlation 
with firing rate helped the cells encode information, provided the correlations were stimulus-dependent. Several theoretical studies have explored this basic structure, and found that it 
is generally beneficial to 
modulate correlations in this way. However --- aside from mouse retinal cells referenced here ---  we do not yet have many examples of real neural circuits that show this correlation-firing rate pattern, 
so we do not know what common features (or mechanisms) might occur between them. 

In this study, we address this question via a computational model.
We set up a computational model with features representative of a generic cortical network, to see whether correlations would increase with firing rate. To produce different firing patterns, we varied excitatory coupling. 
We found that with stronger excitatory coupling, there was
a positive relationship between pairwise correlations and firing rates. 
We used a network linear response theory to show why correlations could increase with firing rates in some networks, but not in others; this could be explained by how cells responded to fluctuations in inhibitory conductances. 


\section*{Introduction}
One prominent goal of modern theoretical neuroscience is to understand how the features of cortical neural networks 
lead to modulation of spiking statistics \cite{van96,renart10,doiron16}. 
This understanding is essential to the larger question of how sensory information is encoded and transmitted, because such statistics are known to
impact 
population coding  \cite{kaybook,dayan2001theoretical,shamir06,josic09,Hu+2014}.
Both experimental and theoretical inquiries are complicated by the fact that neurons are widely known to have heterogeneous attributes \cite{azouz99,padmanabhan10,marsat10,marder06,marder11,Har+2015}.  

One family of statistics that is implicated in nearly all population coding studies is trial-to-trial variability (and co-variability) in spike counts; 
there is now a rich history of studying how these statistics arise, and how they effect coding of stimuli\cite{averbeck06,ecker11,moreno14,ruff14,kohn16}. 
Recent work by numerous authors has demonstrated that the information content of 
spiking neural activity depends on spike count correlations and its relationship (if any) with stimulus tuning \cite{averbeck06,moreno14,silveira14,zylberberg16,kohn16}.  
Since a population of sensory neurons might change their firing rates in different ways to stimuli, 
uncovering the general mechanisms for when spiking correlations increases with firing rate (or when they do not) is important in the 
context of neural coding.  Thus, we study this question in a general recurrent neural network model.

One observation that has been made in some, but not all,  experimental studies is that pairwise correlations increase with firing rates. This relationship has been observed \textit{in vitro} \cite{nature_Rocha_Doiron_ESB} and in several visual areas: area MT \cite{bair01}, V4 \cite{cohen09}, V1 \cite{lin15,schulz2015}, and notably, in ON-OFF directionally sensitive retinal ganglion cells \cite{franke16,zylberberg16}. The retinal studies involved cells with a clearly identified function, and therefore allowed study of the coding consequences of this correlation/firing rate relationship. Both studies found that the \textit{stimulus-dependent} correlation structure observed compared favorably to a structure in which \textit{stimulus-independent} correlations were matched to their (stimulus-)averaged levels. This finding reflects a general principle articulated in other studies \cite{moreno14,kohn16}, that stimulus-dependent correlations are beneficial when they serve to spread the neural response in a direction \textit{orthogonal} to the signal space.


While many studies have illustrated the connection between stimulus-dependent correlation structure and coding, these have (until recently: see \cite{lin15,zylberberg16,franke16}) largely taken the correlation structure as given, leaving open the question of 
how exactly a network 
might produce the hypothesized correlation structure \cite{shamir06,josic09} (see also the theoretical calculations in \cite{zylberberg16, franke16}). Theoretical studies of the \textit{mechanisms} that contribute to correlation distributions
have largely analyzed homogeneous networks (i.e. cells are identical, aside from E/I identity) \cite{renart10,pernice11,trousdale12,doiron16}, which does not allow an exploration of a correlation/firing rate relationship. 
Thus, how correlation coefficients can vary across a population of heterogeneously-tuned neurons is not yet well understood despite its possible implications for coding. 

In this paper we investigated the relationship between correlations and firing rates in
conductance-based leaky integrate-and-fire ({\bf LIF}) neural network models, consisting of excitatory (E) and inhibitory (I) cells that 
are recurrently and randomly coupled.  
We introduced neural heterogeneity by allowing thresholds to vary across the population, which induced a wide range of firing rates,
and induced different firing regimes by varying the strength of recurrent excitation.
We found that with relatively strong excitation, pairwise correlations increased with firing rate.

In theoretical studies, this correlation-firing rate trend has been explained in feed-forward networks driven by common input \cite{nature_Rocha_Doiron_ESB,ESB_PRL_2008,ostojic09}. Here we investigated whether the correlation/firing relationship in recurrent networks can be explained by this theory, but where the source of input correlations is internally generated; i.e., from overlapping projections within the recurrent network.  
We first adapted a network linear response theory, to decompose predicted correlations into contributions from different graph motifs, which are subgraphs which form the building blocks of complex networks \cite{zhao_etal_FCN_2011,pernice11,Hu+2013}.
We found distinct patterns in how motifs contribute to pairwise correlation, between the different spiking regimes: with weak excitation, negative third-order motifs partially cancel positive second-order motifs (as in Pernice et al. \cite{pernice11}), thus diluting a possible  correlation/firing rate relationship, whereas third-order motifs reinforce second-order motifs with stronger excitation.

However, in both regimes second-order motifs --- and specifically \textit{inhibitory common input} --- were still the dominant contributor to overall pairwise correlations.  This allowed us to generalize theory from \cite{nature_Rocha_Doiron_ESB}, and describe pairwise correlations in terms of a single-cell susceptibility function. 
Surprisingly, we found that correlations from inhibitory common input could either increase \textit{or} decrease with firing rate, depending on how cells responded to fluctuations in inhibitory conductances.  

%

We further show that a correlation-firing rate relationship has an important consequence for heterogeneous networks; it can shape low-dimensional structure in the correlation matrix.   Low-dimensional structure --- often modeled with a low-rank approximation to the correlation matrix --- is important because it can be used to improve estimation \cite{yatsenko15} and even to reconstruct full correlation matrices from incomplete data \cite{candes10,Sau+12,bishop2014deterministic}; such structure has been observed in experimental data  \cite{goris14,ecker14,lin15,kanashiro16,CY14} but its origin is not always known. 
We demonstrate in our networks that when correlation co-varies with firing rate, 
 the (E-E) correlation matrix could be accurately 
modeled with a low-rank approximation, and the low-rank projection in this approximation was strongly associated with firing rate. 
Thus we demonstrate that low-rank structure 
can result recurrent activity modulated by single-cell characteristics, 
as well as from a global input or a top-down signal \cite{goris14}.

%

\section*{Results}
We studied asynchronous recurrent networks, where we varied the strength of excitation to get different firing behaviors.
We find that the covariation of correlations with firing rates --- a phenomenon observed in feed-forward networks --- occurs here in one firing regime, but not the other.  This could be explained in terms of how single cells responded to fluctuations in inhibitory conductance.
Finally, we show that when correlations covary with firing rates, the correlation matrix admits a low-rank approximation.  

\subsection*{Asynchronous firing in heterogeneous networks} \label{sec:dist_MC_asyn}
\begin{figure}
\centering
\includegraphics[width=\textwidth]{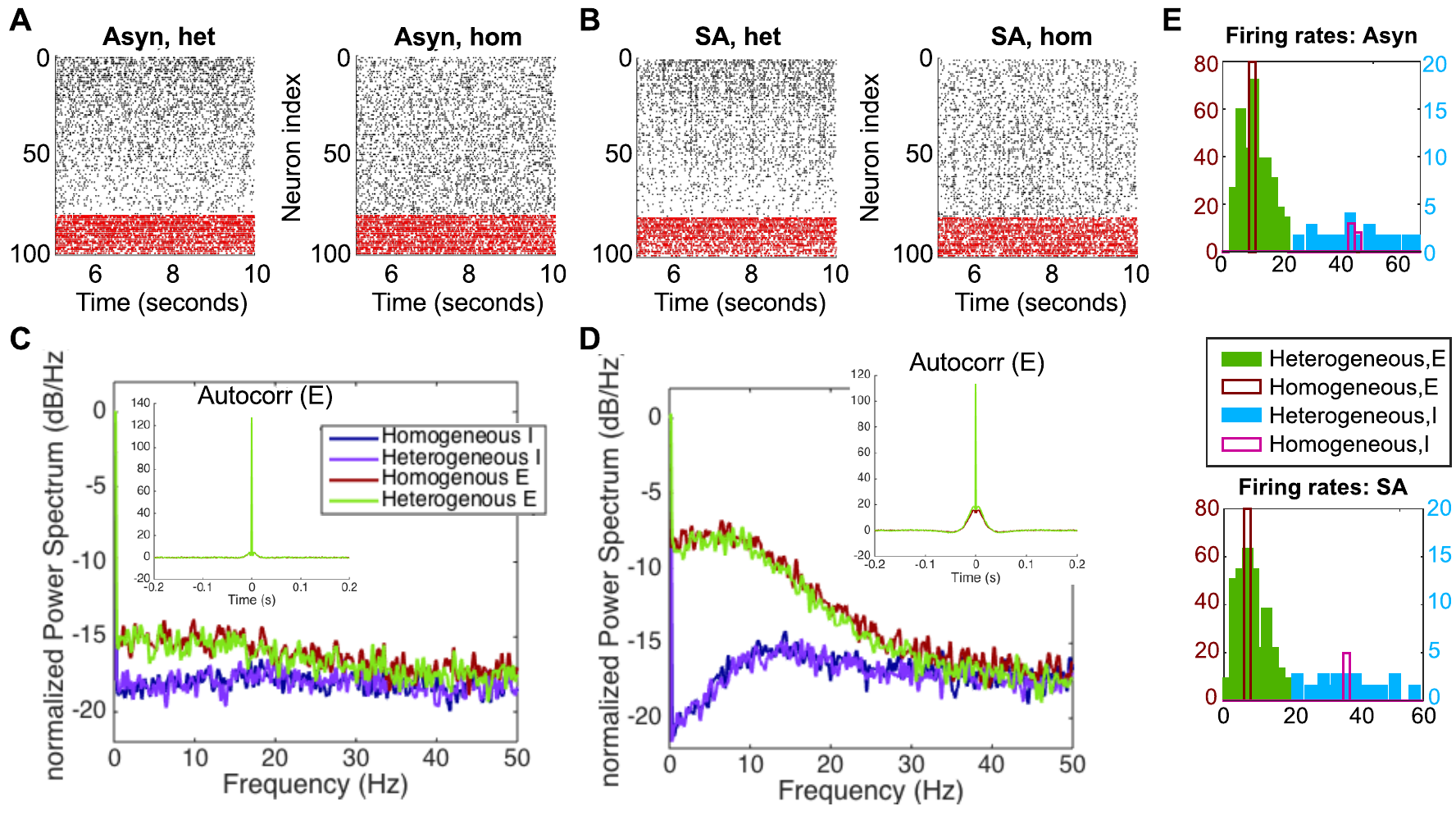}
\caption{\label{fig:tworegimes_fig1} {\bf Two firing regimes in heterogeneous networks.} Monte Carlo simulations illustrating two firing regimes we consider in this paper. 
(A) Raster plots from the asynchronous ({\bf Asyn}) regime.
(B) Raster plots from the \textit{strong asynchronous} ({\bf SA}) regime, showing occasional bursts of activity.
(C) Power spectra in the asynchronous regime.
(D) Power spectra in the strong asynchronous regime.
(E) Firing rates in the asynchronous (top panel) and SA (bottom panel) regimes.
In (A-B), cells are ordered by 
increasing threshold value.  Power spectra (C-D) are normalized to their maximum value and expressed in decibels/Hz. } 
\end{figure}

We performed Monte Carlo simulations of recurrent, randomly connected E/I networks, as described in \ref{sec:meth1}\textbf{Methods: Neuron model and network setup}.  To connect to previous literature on asynchronous spiking, we compared networks with and without single-cell variability ---  referred to as \textit{heterogeneous} and \textit{homogeneous} respectively. Heterogeneity was introduced by allowing cell threshold to vary, which induced a corresponding range of firing rates (see \textbf{Methods: Neuron model and network setup} for details).
We first chose parameters so that the networks exhibited the classical \textit{asynchronous irregular} ({\bf Asyn}) regime, in which each neuron has irregular Poisson-like spiking, correlations are low, and the population power spectra are flat \cite{brunel}. 
In Fig~\ref{fig:tworegimes_fig1}A we show raster plots from both the heterogeneous and homogeneous networks, in this regime. 
The heterogeneous network shows a gradient in its raster plot, because cells are ordered by decreasing firing rate. The population power spectra were flat, for both E and I cells and in both homogeneous and heterogeneous networks (Fig~\ref{fig:tworegimes_fig1}C).

When we increased excitation 
(by increasing both $W_{EE}$ and $W_{IE}$, where $W_{XY}$ is the conductance strength from type $Y$ to $X$; see Table~\ref{table:parameters} for parameter values),  we observed occasional bursts of activity. However, the bursts do not occur at regular intervals and do not involve the entire population (we found excitatory bursts involved at most 25\% of the population). The network is still moderately inhibition-dominated 
and neurons are spiking irregularly; example raster plots are shown in Fig~\ref{fig:tworegimes_fig1}B. 
The population power spectra (Fig~\ref{fig:tworegimes_fig1}D) are no longer flat (compare to the asynchronous regime, Fig~\ref{fig:tworegimes_fig1}C); they show local maxima around 8 Hz, but it is not a pronounced peak.
We will refer to this as the \textit{strong asynchronous} ({\bf SA}) regime \cite{ostojic14}.

In both  Fig~\ref{fig:tworegimes_fig1}C and Fig~\ref{fig:tworegimes_fig1}D, we note that --- despite the apparent differences in the distribution of spikes across the network, evident in the raster plots --- 
both the autocorrelation functions (Fig~\ref{fig:tworegimes_fig1}C,D, insets) and the power spectra from the heterogeneous and homogeneous networks are very similar. 
Thus, we have a fair comparison to examine the role of heterogeneity, independent of other characteristics of the network.

The distribution of both excitatory and inhibitory firing rates are extremely narrow in the homogeneous network, 
but broad in the heterogeneous network (Fig~\ref{fig:tworegimes_fig1}E). This is expected, as each excitatory (inhibitory) cell in the homogenous network has the same uncoupled firing rate; because the number of synaptic inputs is likewise fixed, population variability in synaptic input is limited.  Therefore the heterogeneous networks have a range of firing rates, which allows us to investigate the possibility of a relationship between (variable) firing rate and pairwise correlations.
Population-averaged firing rates were very similar between the heterogeneous and homogeneous networks: in the asynchronous regime $\llangle \nu_E \rrangle=10.6\,$Hz (heterogeneous) and $\llangle \nu_E \rrangle=10.1\,$Hz (homogeneous), while $\llangle \nu_I\rrangle=44.3$\,Hz (heterogeneous) and $\llangle \nu_I\rrangle=43.5$\,Hz (homogeneous). 
 In both regimes Fano factors 
ranged between 0.9 and 1.1, consistent with Poisson-like spiking (more statistics are given in Tables S1 and S3).

\subsection*{Correlation increases with firing rate in the strong asynchronous regime}
%
%

We next sought a possible
relationship between pairwise correlations --- quantified via the Pearson's correlation coefficient for spike counts, $\rho_{ij} \equiv \Cov_T(n_i,n_j)/\sqrt{\Var_T(n_i) \Var_T(n_j)}$ --- and single-cell firing rates.
Such relationships have been found in feed-forward networks 
\cite{nature_Rocha_Doiron_ESB,ESB_PRL_2008,ostojic09}, and 
impact information transfer when considered in concert with stimulus selectivity (i.e. signal correlations) \cite{averbeck06,josic09,Hu+2014,kohn16}. In heterogeneous networks, the large range of firing rates 
--- equivalently the large range of operating points --- admits the possibility that cells at different operating points may differ in their ability to transfer correlations.

\begin{figure}
\centering
 \includegraphics{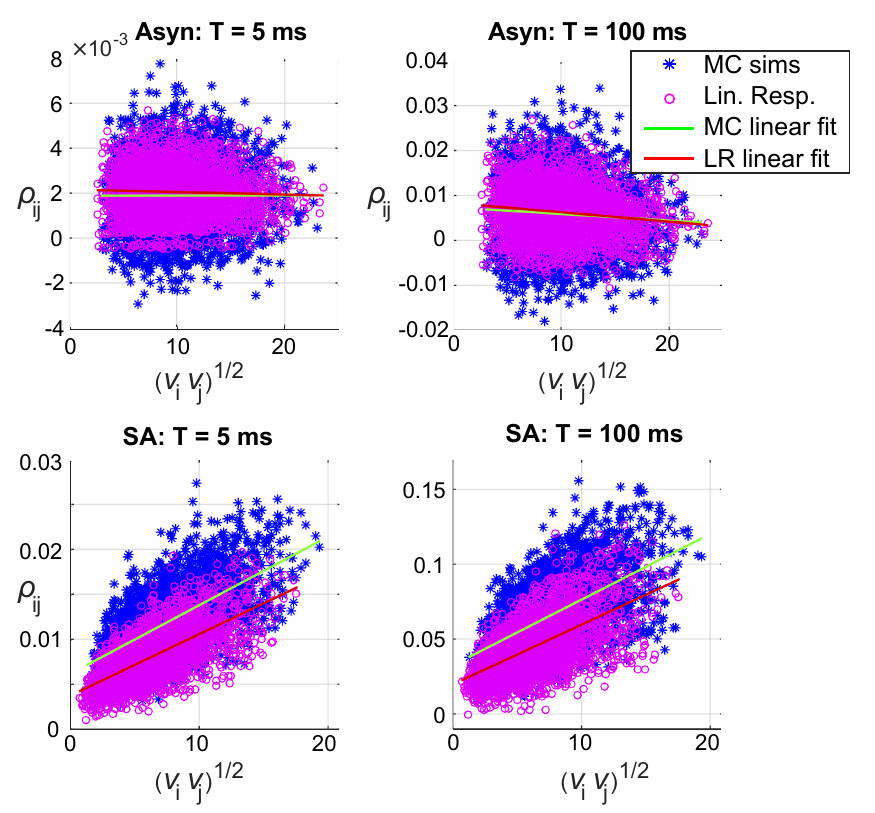}
\caption{\label{fig:rhoEE_vs_fr_MC_and_LR_Asyn_wLM}  {\bf Correlation increases with firing rate in the strong asynchronous regime.} E-E correlation $\rho_{ij}$ vs. geometric mean firing rate $\sqrt{\nu_i \nu_j}$, cell-by-cell comparison of Monte Carlo simulations (blue stars) and linear response (magenta circles), in a heterogeneous network. Left to right: time window $T = $ 5 ms  and 100 ms. Top row: asynchronous regime. Bottom row: strong asynchrony.} 
\end{figure}

To investigate this we plotted pairwise correlations for each distinct excitatory pair $\rho_{ij}$, versus the geometric mean of the firing rates $\sqrt{\nu_i \nu_j}$, in both regimes (asynchronous and strong asynchronous), for a range of time scales (blue stars in Fig~\ref{fig:rhoEE_vs_fr_MC_and_LR_Asyn_wLM}). We focus here on excitatory-excitatory (E-E) pairs, because excitatory synaptic connections provide the predominant means of propagating cortical sensory information to higher layers.
Our results show a striking difference between the two spiking regimes; while there is no clear relationship with firing rate in the asynchronous regime 
(Fig~\ref{fig:rhoEE_vs_fr_MC_and_LR_Asyn_wLM}, top row), the 
strong asynchronous regime shows a distinct positive trend with firing rate (Fig~\ref{fig:rhoEE_vs_fr_MC_and_LR_Asyn_wLM}, bottom row). 
We can quantify a hypothesized relationship between $\nu$ and $\rho$ 
with linear regression, and indeed find that geometric mean firing rate explains a substantial part of the variability of correlations in the strong asynchronous regime obtained from the Monte Carlo simulations, 
with $R^2$ values (i.e. percentage of variability explained) of 
0.41, 0.37, and 0.34 for time 
windows of $T=5$, $50$, and $100$ ms respectively (in contrast, $R^2$ values for the asynchronous network are below 0.005). 

In recurrent networks, the response of each cell is shaped by both direct and indirect connections through the network. 
We used the linear response theory described in 
{\bf Methods: Linear Response Theory} and {\bf Methods: Computing statistics from linear response theory}
to predict the full correlation matrix $\Cvec_T$ at various time scales, including the limit of long time scales: $\tilde{\Cvec}(0) = \lim_{T \rightarrow \infty} \frac{1}{T} \Cvec_T$.  We found that this theory successfully captured E-E correlations, both the full distribution of values and coefficients of individual cell pairs (details, including figures, can be found in: \textbf{Supporting Information: S1 Text}). 

We then plotted the predicted correlation, $\tilde{\Cvec}_{ij}/\sqrt{\tilde{\Cvec}_{ii}\tilde{\Cvec}_{jj}}$, vs. geometric mean firing rate $\sqrt{\nu_i \nu_j}$ (magenta circles in 
Fig~\ref{fig:rhoEE_vs_fr_MC_and_LR_Asyn_wLM}). The predicted correlations captured the same positive relationship observed in Monte Carlo results, with $R^2$ values of 0.47, 0.4, and 0.36.

\subsection*{Decomposition of correlation by graph motifs shows a distinct pattern for each spiking regime}

\begin{figure}
\centering
 \includegraphics[width=0.7\textwidth]{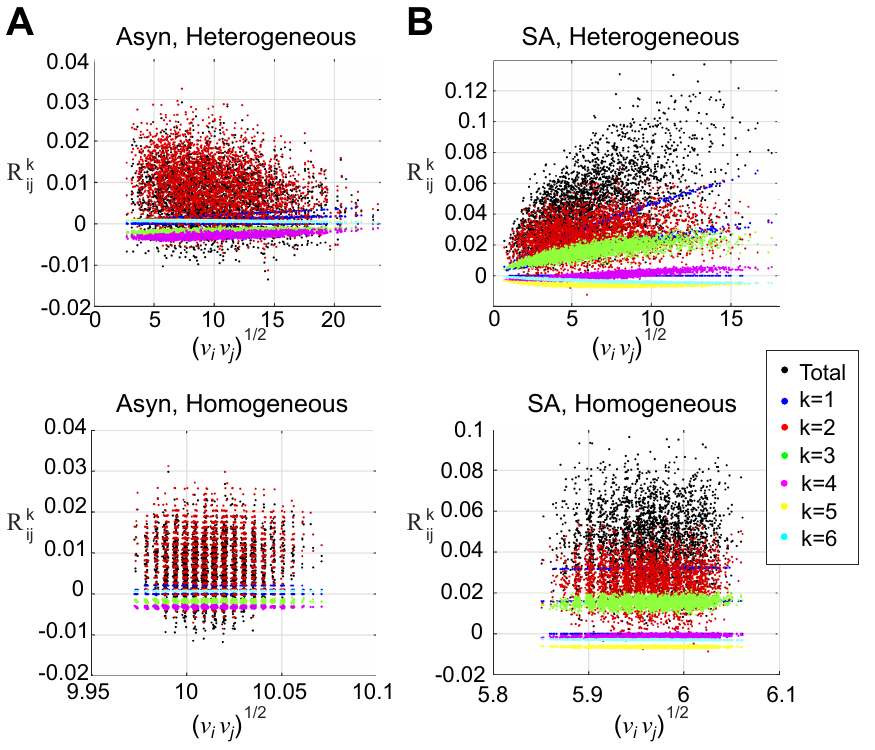}
\caption{\label{fig:Corr_DiffOrders_vs_FR} {\bf Pairwise correlations are built from graph motifs.} 
Contributions of different orders to prediction of E-E correlations in the asynchronous (A) and strong asynchronous (B) regimes with linear-response theory. 
Both heterogeneous (top panels) and homogenous (bottom panels) networks are shown.  
(A) First order contributions (blue dots) have a 1-1 correspondence (separated into three curves for three possible first order connectivity structures).  Second order contributions (red) are positive, third order (green) are negative 
leading to \textit{cancellation}; fourth (magenta) order contributions are negative while fifth and sixth are miniscule.  
(B) First and second order contributions are similar to (A) but with a stronger dependance on firing rate. Third order contributions (green) are distinctly positive so that there is \textit{not} cancellation as in (A).  
See main text for further discussion.  
Bottom panels of (A,B) show results from homogeneous networks. The magnitude and range of each term is similar as in the corresponding heterogeneous network (top panel), but with less variation in firing rate.  
} 
\end{figure}

Why does a correlation/firing rate relationship emerge in one spiking regime, but not the other? In feed-forward networks, a positive correlation/firing rate relationship results from transferring common input through fluctuation-driven, 
asynchronously-firing cells \cite{nature_Rocha_Doiron_ESB,ESB_PRL_2008}.  
In contrast, the amount of shared input into two cells in a recurrent network is determined by both direct and indirect connections through the network. To separate the impact of different network pathways, 
we decomposed the \textit{linear response-predicted} 
correlations at long time scales (i.e. $\tilde{\Cvec}(0) = \lim_{T \rightarrow \infty} \frac{1}{T} \Cvec_T$) into normalized contributions from $n$-th order motifs, as described in 
{\bf Methods: Quantifying the role of motifs in networks}. Common input from a divergent connection, for example, results from the 2nd-order motif $\Kvec^{\ast} \Cvec^0 \Kvec$.
In Fig~\ref{fig:Corr_DiffOrders_vs_FR}, we plot the summed contributions up to sixth order --- i.e. $\tilde{\Rvec}_{ij}^k$, for $k=1,2,...6$
--- versus geometric mean firing rate, $\sqrt{\nu_i \nu_j}$. The total normalized correlation, $\tilde{\Cvec}_{ij}/\sqrt{\tilde{\Cvec}_{ii} \tilde{\Cvec}_{jj}}$, is shown as well. In all cases, we plot long time scale correlations $\omega = 0$; 
each distinct E-E pair is represented.  

In the asynchronous regime (top panel of Fig~\ref{fig:Corr_DiffOrders_vs_FR}A),
first-order contributions ($\tilde{\Rvec}^1$) separate into  three distinct ``curves", reflecting a 1-1 relationship with firing rate conditioned on first-order connectivity (no connection between $i$ and $j$; one connection between $i$ and $j$; bidirectional connection between $i$ and $j$). This might be expected if susceptibility to excitatory conductances, $\tilde{A}_{g_E}(0)$, varies roughly like firing rate (as we will argue later); thus, the contribution from an $i \rightarrow j$ connection should approximately vary like: $\nu_i \tilde{A}_{g_E, j} (0)/\sqrt{\nu_i \nu_j}  \propto \nu_i  \nu_j /\sqrt{\nu_i \nu_j} = \sqrt{\nu_i \nu_j}$.
Second-order contributions are overall positive while third-order contributions are overall negative (consistent with \cite{pernice11});  neither appear to have a relationship with firing rate.
Second-order contributions are conspicuously dominant; fifth and sixth order terms are near zero.

It is instructive to compare the results from the homogeneous network, shown in the bottom panel.  Since firing rate is nearly uniform across the network, we would not expect to see a relationship; rather, we might use this to gauge the relative magnitude of contributions at different orders, independent of the complicating effect of firing rate.   
Indeed, we see similar pattens as in the heterogeneous network; $\tilde{\Rvec}_{ij}^1$ breaks into three distinct curves and is positive; $\tilde{\Rvec}_{ij}^2$ is generally positive and $\tilde{\Rvec}_{ij}^3$ generally negative; the magnitude of $\tilde{\Rvec}_{ij}^2$ overshadows both $\tilde{\Rvec}_{ij}^1$ and $\tilde{\Rvec}_{ij}^3$.

This qualitative picture changes when we consider the strong asynchronous regime, in Fig~\ref{fig:Corr_DiffOrders_vs_FR}B. 
First-order contributions follow a similar pattern as in the asynchronous regime, and second-order contributions are likewise positive.
However, third-order contributions are positive, and in the heterogeneous network they have a distinctly positive relationship with firing rate (top panel). 
Thus, in the asynchronous regime, negative third-order contributions partially cancel with positive second-order contributions;
in the strong asynchronous regime,  
first, second, and third-order motifs reinforce each other, contributing to an overall positive relationship with firing rate (black dots).

\begin{figure}
\centering
\includegraphics{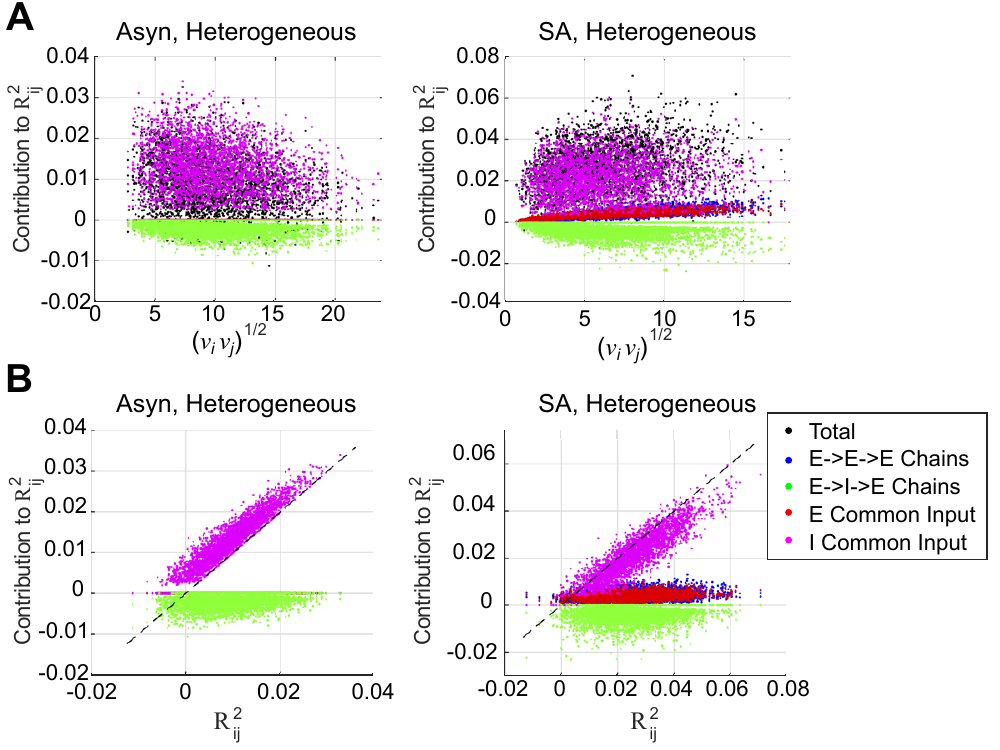}
\caption{\label{fig:Corr_DiffMotif_2ndOrders}  {\bf Inhibitory common input is the dominant second-order motif in both asynchronous and strong asynchronous networks} (A) Contributions of different 2nd-order motifs to prediction of E-E correlations in a heterogeneous network, in the asynchronous (left) and strong asynchronous (right) regimes. Left: inhibitory common input (magenta) is dominant but is partially cancelled by $E\rightarrow I\rightarrow E$ chains (green); common $E$ inputs (red) and $E\rightarrow E \rightarrow E$ chains (blue) are near 0.  
Right: In contrast to the asynchronous regime, common $E$ inputs (red) and $E\rightarrow E\rightarrow E$ chains (blue) are positive and vary with firing rate, preventing dilution of correlation.  
(B) As in (A), but plotted vs. total contribution from second-order motifs $\tilde{\Rvec}^2$. The left panel shows that inhibitory common input (magenta) is balanced by $E\rightarrow I\rightarrow E$ chains (green), because it is above the unity line. 
The right panel shows that the inhibitory common input (magenta) is instead reinforced by other motifs, because it is below the unity line.} 
\end{figure}

Despite these differences, second-order contributions appear to dominate both regimes. Therefore, we next analyze contributions from specific \textbf{second-order motifs} in 
Fig~\ref{fig:Corr_DiffMotif_2ndOrders}. There are four distinct second-order motifs that can correlate two E cells. There are two types of chains, from $\Kvec^2 \Cvec^0$ and $\Cvec^0 \left( \Kvec^* \right)^{2}$.  
An $E\rightarrow E\rightarrow E$ chain tends to positively correlate; an $E \rightarrow I \rightarrow E$ chain will negatively correlate; these are shown as blue and green respectively.
There are two types of common input, from $\Kvec \Cvec^0 \left( \Kvec^* \right)$; they correspond to common input from E and I cells, i.e. $E \leftarrow E \rightarrow E$ and $E \leftarrow I \rightarrow E$. 
They \textit{both} lead to positive correlations and are shown as red and magenta respectively.

In the asynchronous regime (left panel of Fig~\ref{fig:Corr_DiffMotif_2ndOrders}A), the dominant contributions are $I$ common input (magenta) and negative ($E \rightarrow I \rightarrow E$) chains (green); correlating chains (blue) and excitatory common input (red) are barely visible, as they are clustered near zero. 
This contrasts with the strong asynchronous case (right panel): blue and red dots are now visible, similar in magnitude, and 
show a clear 1-1 trend with firing rate. As a result, they prevent ``dilution" of correlation from the decorrelating chains (green dots). In both regimes, \textit{inhibitory common input} appears to be the dominant second-order motif. In Fig~\ref{fig:Corr_DiffMotif_2ndOrders}B we plot the contribution from different second-order motifs vs. the total contribution from second-order motifs, $\tilde{\Rvec}_{ij}^2$ (rather than geometric mean firing rate, $\sqrt{\nu_i \nu_j}$). In both panels, the inhibitory common input (magenta) clusters around the unity line, showing it is the best predictor of the total second-order contribution.
%

In conclusion, decomposition of pairwise correlations into graph motifs has shown us two important things: first, while third-order motifs contribute to the positive correlation/firing rate relationship observed in the SA regime, second-order motifs still dominate in both regimes. Second, {\bf inhibitory common input is the most important second-order motif in both regimes} 
(Fig~\ref{fig:Corr_DiffMotif_2ndOrders}B). 

\subsection*{Susceptibility to inhibition can either increase or decrease with firing rate }
In feedforward networks --- i.e. in the absence of a path between two cells --- correlations in \textit{outputs} (i.e. spike trains) must arise from correlations in \textit{inputs}; for example, through shared or common inputs. We have found that inhibitory common input is the dominant contributor to pairwise correlations in both the asynchronous and strong asynchronous regimes; we now turn our attention to modeling this term (inhibitory common input) specifically. 

Previous work that analyzed the relationship between the long-time correlation and firing rate in feedforward networks \cite{nature_Rocha_Doiron_ESB,ESB_PRL_2008} quantified a \textit{susceptibility} function that measures the ratio between output and input correlations:
\begin{eqnarray}
S & \approx & \frac{\rho}{c}. 
\end{eqnarray}
If both cells receive a large (but equal) number of uncorrelated inputs, $c$ would be the fraction of inputs that are common to both $i$ and $j$. 

In the networks examined here, each cell had a fixed in-degree for both excitatory and inhibitory cells;  however, for any given \textit{pair} of cells $i$ and $j$, the number of E and I inputs that synapsed 
onto both cells will vary from pair to pair. Thus, we next considered the possibility that our (negative) 
finding in the asynchronous network could be explained by accounting for variable $c_{ij}$. 

\begin{figure}
\centering
  \includegraphics[width=\textwidth]{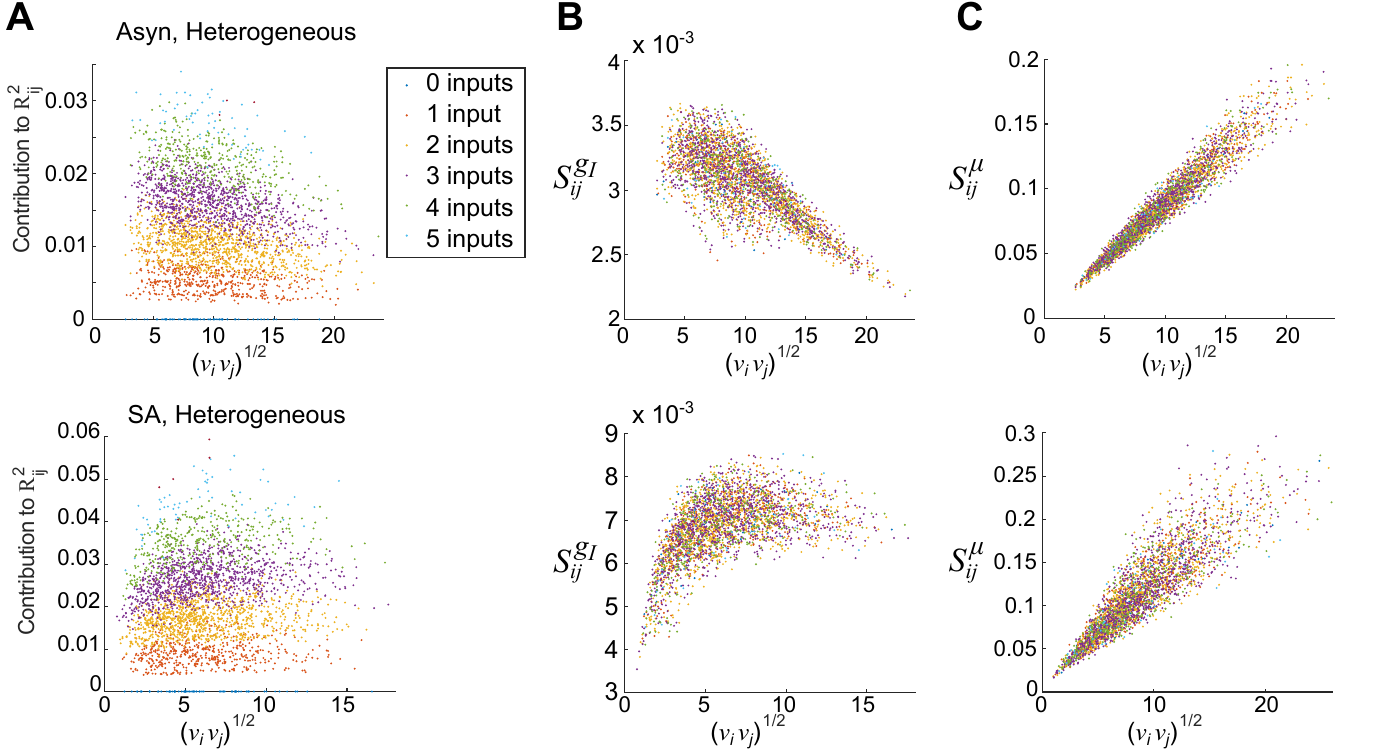}
\caption{\label{fig:Corr_vs_FR_Het_by_NComInp} 
{\bf Susceptibility to conductance fluctuations can explain correlation-firing rate relationships.} In (A-C): heterogeneous asynchronous (top) and heterogeneous strong asynchronous (bottom). (A) Correlation ($\rho$) from I common inputs vs. firing rate, segregated based on the number of common inhibitory inputs. 
(B) Estimated correlation susceptibility to fluctuations in inhibitory \textit{conductances} vs. firing rate ($S^{g_I}_{ij}$ ).
(C) Correlation susceptibility to fluctuations in inhibitory \textit{currents} vs. firing rate ($S^{\mu}_{ij}$).
} 
\end{figure}

We focus on inhibitory common input, which is the dominant second-order contribution in the asynchronous network (Fig~\ref{fig:Corr_DiffMotif_2ndOrders}). We segregated pairs by whether they had $0$, $1$, $2$, etc.. common inhibitory inputs; we then use this number as a proxy for $c$ (recall that each excitatory cell had exactly 7 inhibitory inputs, so that this number divided by $7$ approximates the common input fraction; two common inputs imply $c \approx 0.28$ for example).  We plot the results for the asynchronous network in Fig~\ref{fig:Corr_vs_FR_Het_by_NComInp}A, top panel (data for each distinct value of $c$ is presented by color). As we might expect, correlation increases as $c$ increases.  However, for a fixed $c$,  
there is not an apparent relationship between firing rate and correlation; if anything, there appears to be a slight decrease.  
Correlation also increases with $c$ in the strong asynchronous network (Fig~\ref{fig:Corr_vs_FR_Het_by_NComInp}A, bottom panel); however, here we also see a modest increase with geometric mean firing rate $\sqrt{\nu_i \nu_j}$. 

Previous theoretical work \cite{nature_Rocha_Doiron_ESB,ESB_PRL_2008} identified an increase in susceptibility with firing rates in \textit{current-driven} neurons; we next considered the possibility that this fails to hold for conductance-driven neurons. As described in {\bf Methods: Quantifying correlation susceptibility}, we estimated correlation susceptibility for each pair of neurons, by using the susceptibility function for each neuron to conductance fluctuations (computed as part of the linear response theory), divided by a measure of the long-timescale spike count variance:
\begin{eqnarray}
S^{\langle g_I \rangle}_{ij} & = & \frac{\tilde{A}_{\langle g_I \rangle, i}(0) \tilde{A}_{\langle g_I \rangle, j}(0)}{\sqrt{\tilde{\Cvec}_{ii}(0)\tilde{\Cvec}_{jj}(0)}}
\end{eqnarray}
We plotted the results for both networks in Fig~\ref{fig:Corr_vs_FR_Het_by_NComInp}B; while susceptibility increases with firing rate in the strong asynchronous network (except for the largest firing rates), it actually decreases with firing rate in the asynchronous network. 

We can contrast with the estimated susceptibility to \textit{current} fluctuations (i.e. $A_{\mu,i}$, with $\mu_i$, $\tau_{{\rm eff},i}$, and $\sigma_{{\rm eff},i}$ as in 
Eq~\ref{eqn:eff_for_cond_based}) which we also computed for the same set of cell pairs, shown in Fig~\ref{fig:Corr_vs_FR_Het_by_NComInp}C.  
\begin{eqnarray}
S^{\mu}_{ij} & = & \frac{\tilde{A}_{\mu, i}(0) \tilde{A}_{\mu, j}(0)}{\sqrt{\tilde{\Cvec}_{ii}(0)\tilde{\Cvec}_{jj}(0)}}
\end{eqnarray}
Here, we see that $S^{\mu}_{ij}$ increases with firing rates, in both networks. 

To understand the difference between $S^{g_I}_{ij}$ and $S^{\mu}_{ij}$, we consider how the cell responds differently to currents vs. conductances. Both networks are inhibition-dominated, with resting potentials in the range $(-0.33,-0.28)$ (asynchronous)  and $(-0.183,-0.075)$ (strong asynchronous); since the cells are operating near the inhibitory reversal potential ($\mathcal{E}_I = -0.5$), fluctuations in the inhibitory conductance have limited effect on the firing rate of the cell. This effect is more pronounced in the asynchronous regime, where the small distance between the resting and synaptic potentials creates a saturating effect on the firing rate; previous authors found susceptibility decreases with firing rate, in a simple thresholding model in which the f-I curve saturates\cite{nature_Rocha_Doiron_ESB}. In contrast, susceptibility to \textit{excitatory} conductances, $S^{g_E}_{ij}$, more closely resembled the current input susceptibility $S^{\mu}_{ij}$. 
The cells are operating far from the excitatory reversal potential ($\mathcal{E}_E = 6.5$); therefore fluctuations in $g_E$ are multiplied by a relatively constant $\mathcal{E}_{E}-V$ and thus have a ``current-like" effect. 

\begin{figure}
\centering
 \includegraphics[width=\textwidth]{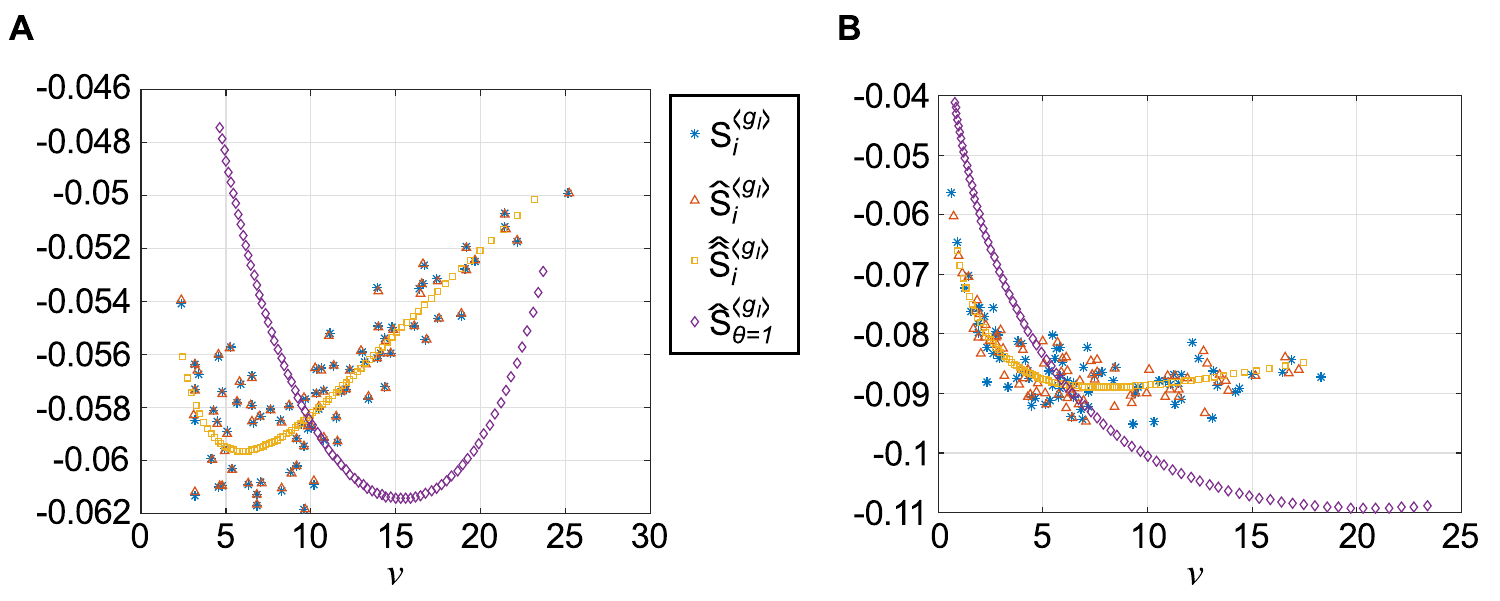}
\caption{{\bf How firing rate diversity is achieved in a heterogeneous network will affect susceptibility.} Single-cell susceptibility function(s) for a conductance-based LIF neuron, as a function of firing rate $\nu$. 
Successive approximations shown are: original single-cell susceptibility, $S^{\langle g_I \rangle}_{i}$ (Eq~\ref{eqn:susc_single_cell}, blue stars); most parameters set to average value, $\hat{S}^{\langle g_I 
\rangle}_{i}$ (Eq~\ref{eqn:susc_fix_param}, red triangles); all parameters but $\theta_i$ set to average value, $\hat{\hat{S}}^{\langle g_I \rangle}_{i}$ (Eq~\ref{eqn:susc_fix_ALL_param}, gold squares); and $
\theta$ fixed, $\hat{S}^{\langle g_I \rangle}_{\theta=1}$ (Eq~\ref{eqn:susc_fix_theta}, purple diamonds). (A) Asynchronous regime. (B) Strong asynchronous regime.} \label{fig:diversity_mech_susc}
\end{figure}

However, the difference between a threshold-linear vs. saturating f-I curve cannot be a perfect analogy, because for our neurons firing rate and susceptibility are not functions of a single parameter $\mu$: instead, they can depend on all \textit{six} parameters defining the cell; specifically, we define the single-cell susceptibility
\begin{eqnarray}
S^{\langle g_I \rangle}_{i} &\equiv& \frac{\tilde{A}_{\langle g_I \rangle,i}(0)}{\sqrt{\nu_i}}    \label{eqn:susc_single_cell}
\end{eqnarray}
where
\begin{eqnarray} 
\nu_i & = & f \left( \langle g_{I,i} \rangle, \sigma_{I,i}, \langle g_{E,i} \rangle, \sigma_{E,i}, \sigma_i, \theta_i \right) \\
\tilde{A}_{\langle g_I \rangle,i}(0) & = & \frac{\partial f}{\partial x_1}\left( \langle g_{I,i} \rangle, \sigma_{I,i}, \langle g_{E,i} \rangle, \sigma_{E,i}, \sigma_i, \theta_i \right) 
\end{eqnarray}
This quantity is shown in Fig~\ref{fig:diversity_mech_susc}, where it is plotted vs. firing rate $\nu_i$ (blue stars). 
Note that this is a negative quantity; since the susceptibility for a neuron pair $S^{\langle g_I \rangle}_{ij} = S^{\langle g_I \rangle}_{i} S^{\langle g_I \rangle}_{j}$ is 
the product (and therefore positive), 
an \textit{increase} in $S^{\langle g_I \rangle}_{i}$ will result in a \textit{decrease} in $S^{\langle g_I \rangle}_{ij}$ and vice versa.  
We have also used the asynchronous spiking assumption, that $\tilde{\Cvec}_{ii}(0) \approx \nu_i$.

\begin{figure}
\centering
\includegraphics[width=\textwidth]{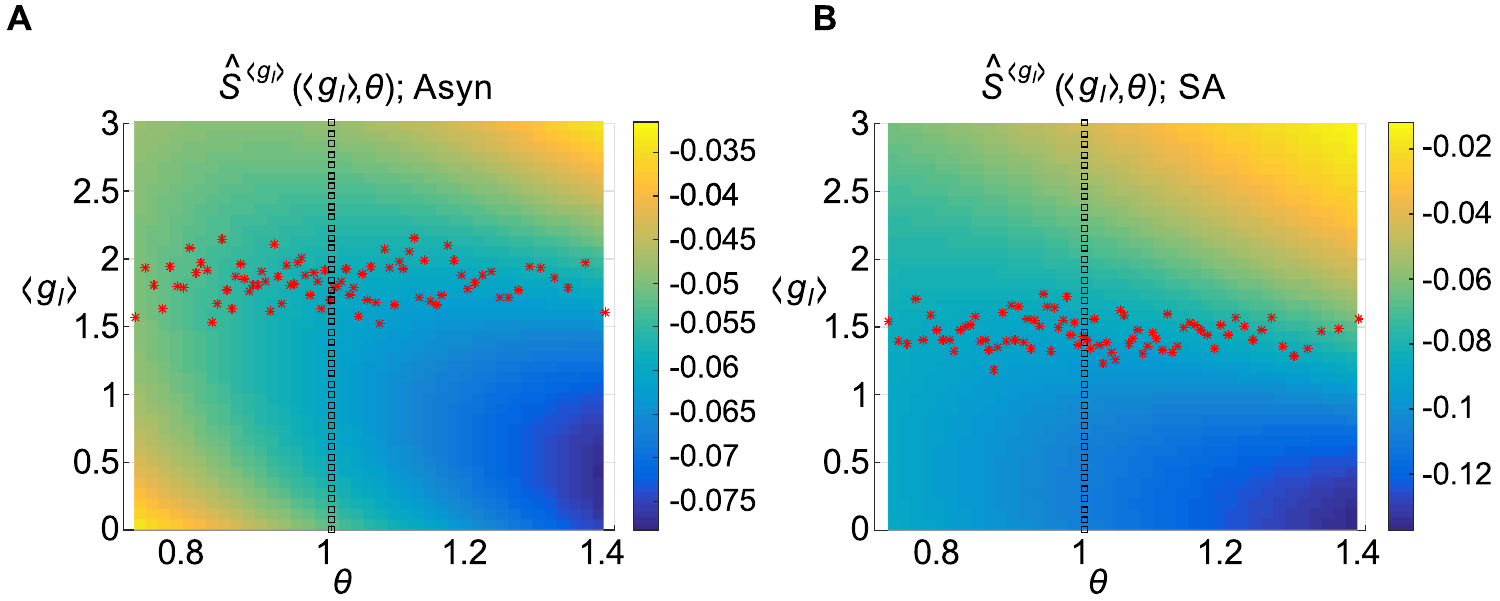}
\caption{{\bf Susceptibility as a function of inhibitory conductance and threshold.} Single-cell susceptibility function for a conductance-based LIF neuron, as a function of mean inhibitory conductance $\langle 
g_I \rangle$ and threshold $\theta$: $\hat{S}^{\langle g_I \rangle} (\langle g_I \rangle, \theta)$ (defined in Eq~ \ref{eqn:susc_fix_param}). 
Other parameters are set to the population average.  Overlays show $(\langle g_{I,i} \rangle, \theta_i)$ values of the actual cells in the network (red stars) and an alternative curve through the plane, $(\langle 
g_{I} \rangle, 1)$, along which comparable firing rate diversity can be observed (black squares). (A) Asynchronous regime. (B) Strong asynchronous regime.} \label{fig:diversity_mech_susc_2D}
\end{figure}

By examining the dependence of susceptibility on its parameters, we hypothesized that it was most important to capture dependence on mean inhibitory conductance and threshold (see \textbf{S1 Text: Approximating single-cell susceptibility in a heterogeneous network}).
We approximated $S^{\langle g_I \rangle}_{i}$, by reevaluating the firing rate function in which $\sigma_{I,i}$, $\langle g_{E,i} \rangle$, $\sigma_{E,i}$ and  $\sigma_i$ have been replaced by their average values: i.e. 
\begin{eqnarray}
\hat{S}^{\langle g_I \rangle}_{i} & \equiv &\frac{1}{\sqrt{F( \langle g_{I,i} \rangle, \theta_i)} }  \frac{\partial F}{\partial x_1}\left( \langle g_{I,i} \rangle, \theta_i \right)   \label{eqn:susc_fix_param}
\end{eqnarray}
 where
 \begin{eqnarray}
F( \langle g_{I,i} \rangle, \theta_i) & \equiv &  f \left( \langle g_{I,i} \rangle, \langle \sigma_{I,i} \rangle_p, \langle \, \langle g_{E,i} \rangle \, \rangle_p, \langle \sigma_{E,i} \rangle_p, \langle \sigma_i \rangle_p, \theta_i \right)
\end{eqnarray}
and  $\langle \, \cdot \, \rangle_p$ denotes the population average.  
The results are also illustrated in Fig~\ref{fig:diversity_mech_susc} (red triangles). In the asynchronous regime (Fig~\ref{fig:diversity_mech_susc}A), the results are remarkably close to the original quantities, indicating that using average parameter values has little effect; in the strong asynchronous regime (Fig~\ref{fig:diversity_mech_susc}B) the difference is larger, but the points appear to occupy the same ``cloud".  However, we can now visualize the susceptibility as a function of only \textit{two} parameters, and we do so in Fig~\ref{fig:diversity_mech_susc_2D} by evaluating $ \hat{S}^{\langle g_I \rangle}_{i}$ on a $(\theta, \langle g_I \rangle)$ grid; the points corresponding to the actual excitatory cells in our network are illustrated in red. In both the asynchronous and strong asynchronous regimes, the red stars form a scattered cloud around the average value $\langle \, \langle g_{I,i} \rangle \, \rangle_p$, with no obvious relationship with $\theta_i$. 

This fact motivated a further simplification, 
\begin{eqnarray}
\hat{\hat{S}}^{\langle g_I \rangle}_{i} &\equiv& \frac{1}{\sqrt{F( \langle \, \langle g_{I,i} \rangle \rangle_p, \theta_i)} }  \frac{\partial F}{\partial x_1}\left( \langle \, \langle g_{I,i} \rangle \, \rangle_p, \theta_i \right)  \label{eqn:susc_fix_ALL_param}
 \end{eqnarray}
 i.e., we replaced $\langle g_{I,i} \rangle$ with its population average,  $\langle \, \langle g_{I,i} \rangle \, \rangle_p$.  The results are shown in Fig~\ref{fig:diversity_mech_susc} (gold squares) and (as we should expect) allow us to discern a one-to-one relationship with firing rate $\nu_i$; importantly, it appears to capture the average behavior of the actual susceptibility values $S^{\langle g_I \rangle}_{i}$. Here, we can see clearly that in the asynchronous regime, correlations should actually \textit{decrease} with firing rate, for $\nu_i > 5$ Hz. In the strong asynchronous regime, correlations will \textit{increase} with firing rate, saturating around 10-15 Hz.

Finally, recall that our actual network sampled a relatively small part of the $(\theta, \langle g_I \rangle)$ plane. This may be attributed to the fact that we generated firing rate diversity (and therefore heterogeneity), by modulating cell excitability through the cell threshold $\theta_i$.  How might our results have changed, if 
we had generated firing rate diversity through some other mechanism? In both regimes, we can increase firing rates by either decreasing $\langle g_{I,i} \rangle$, or by decreasing $\theta$ (see Fig. S7). To explore this, we computed susceptibility values along another curve in the $(\theta, \langle g_I \rangle)$ plane; specifically, we held $\theta$ fixed and instead varied $\langle g_I \rangle$ (illustrated with black squares on Fig~\ref{fig:diversity_mech_susc_2D}); i.e.

\begin{eqnarray} 
\hat{S}^{\langle g_I \rangle}_{\theta=1} (\langle g_I \rangle) &\equiv& \frac{1}{\sqrt{G( \langle g_{I} \rangle, \theta)} } \frac{\partial G}{\partial x_1}\left(  \langle g_{I} \rangle, \theta \right)  \bigg|_{\theta = 1}    \label{eqn:susc_fix_theta}
\end{eqnarray}
where  \[ \qquad G( \langle g_{I} \rangle, \theta)  = f \left( \langle g_{I} \rangle, \langle \sigma_{I,i} \rangle_p, \langle \, \langle g_{E,i} \rangle \, \rangle_p, \langle \sigma_{E,i} \rangle_p, \langle \sigma_i \rangle_p, \theta \right)
\]
Results are shown in Fig~\ref{fig:diversity_mech_susc} (purple diamonds) and show a strikingly different relationship with firing rate; in the asynchronous regime, correlations should increase with firing rate for $\nu < 15$ Hz; in the strong asynchronous regime correlations will increase with firing rate, saturating near 20 Hz.

\subsection*{Low-rank structure in neural correlations is mediated by correlation-firing rate relationship}
Previous work has identified low-dimensional structure in neural correlation matrices \cite{goris14,ecker14,lin15,kanashiro16,CY14};
its origin is not always known \cite{doiron16}.
We next hypothesized that the positive correlation-firing rate relationship we observed in the strong asynchronous regime, might be reflected in low-dimensional structure in the correlation matrix. For simplicity, suppose that correlations were really represented by a function of firing rate (as in \cite{nature_Rocha_Doiron_ESB}): i.e. 
$\rho_{ij} = c S(\nu_i) S(\nu_j)$.  Then we could represent the \textit{off-diagonal} part of the correlation matrix as $\Cvec_T = c \, \Svec \, \Svec^T$, where $\Svec$ is a length $N$ vector such that $\Svec_i = S(\nu_i)$; that is, $\Cvec_T$ would be a rank-one matrix.
%

We followed the procedure outlined in \textbf{Methods: Low-rank approximation to the correlation matrix}
to approximate each correlation matrix, $\Cvec_T$, as the sum of a diagonal matrix and low-rank matrix:
\begin{eqnarray}
\Cvec_T \,  \approx \, \Cvec_T^{\rm diag +R1}& = & \lambda \Ivec + (\sigma_1-\lambda) \uvec_1 \uvec_1^T
\end{eqnarray}
where $\lambda$ is given in closed form by the eigenvalues of $\Cvec_T$:
\begin{eqnarray}
\lambda & = & \lambda_1 - \frac{\sum_{j>1} (\lambda_1 - \lambda_j)^2}{\sum_{j>1} \lambda_1 - \lambda_j}  \label{eqn:lambda_rankone}
\end{eqnarray}
and $\sigma_1$, $\uvec_1$ are the first singular value and singular vector of $\Cvec_T$.
\begin{figure}
\centering
   \includegraphics{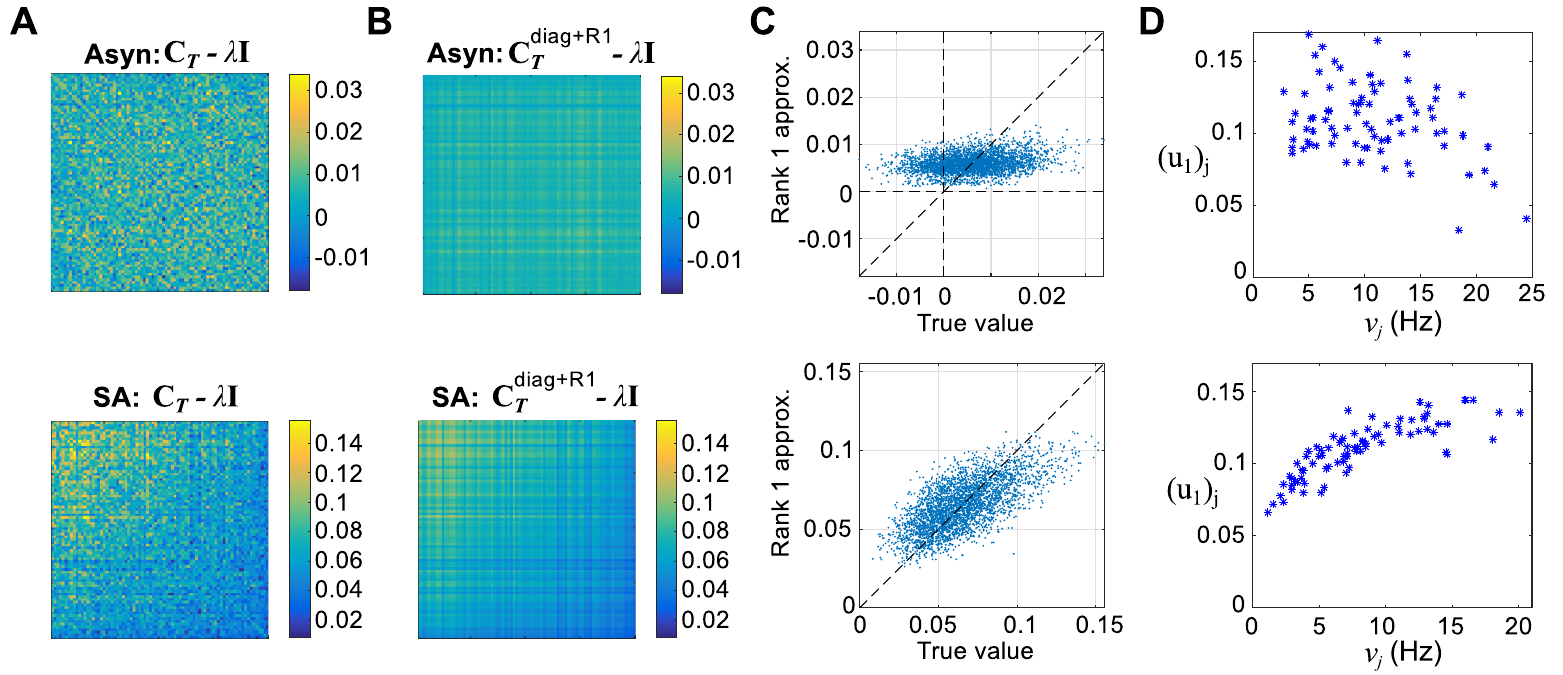}
\caption{\label{fig:lowRank_Asyn_vs_Str_Het} {\bf Low-rank structure in correlation matrices.} Approximating correlation matrices for the heterogeneous networks as a diagonal plus rank-one.  In each column 
of (A-D), the asynchronous (top) and strong asynchronous (bottom) regimes are shown; $T = $ 100 ms.(A)  The shifted E-E correlation matrix, $\Cvec_T-\lambda \Ivec$, for an appropriately chosen $\lambda$. 
(B) A rank-one approximation to $\Cvec_T-\lambda \Ivec$. (C) True correlation coefficients vs. rank-one approximation, cell-by-cell. (D) Weight in the first singular vector, $\uvec_1$ vs. 
geometric mean firing rate $\sqrt{\nu_i \nu_j}$.} 
\end{figure}

In Fig~\ref{fig:lowRank_Asyn_vs_Str_Het}, we show the results from heterogeneous networks in both the asynchronous (top panel in each subfigure) and strong asynchronous (bottom panel in each subfigure) regimes. We first show $\Cvec_T  - \lambda \Ivec$, where $\lambda$ is given by Eq~\ref{eqn:lambda_rankone}, in Fig~\ref{fig:lowRank_Asyn_vs_Str_Het}A. Cells are ordered by (decreasing) firing rate. While no pattern is visible in the asynchronous state (top panel), the strong asynchronous state (bottom panel) shows larger values in the upper left corner, suggesting that correlation increases with firing rate.
This is even more visible in the rank one approximation, $(\sigma_1-\lambda) \uvec_1 \uvec_1^T$, shown in  Fig~\ref{fig:lowRank_Asyn_vs_Str_Het}B. 

We now use $\Cvec_T^{\rm diag + R1}$ to approximate $\Cvec_T$, and compare the results, cell pair-by-cell pair (Fig~\ref{fig:lowRank_Asyn_vs_Str_Het}C). In the asynchronous network, the approximated correlations take on a narrow range (between $0$ and $0.01$, compared to between $-0.015$ and $0.03$ for the measured coefficients) and do not show an obvious positive relationship. In the strong asynchronous regime, the range is more accurate ( between $0.02$ and $0.1$, vs. $0.01$ and $0.15$ for the measured coefficients) and the points cluster around the unity line. 

In Fig~\ref{fig:lowRank_Asyn_vs_Str_Het}D, we plot the weight of each cell in the first singular vector, $(\uvec_1)_j$ vs. the firing rate $\nu_j$. We can clearly see a positive relationship in the strong asynchronous regime (bottom panel), suggesting that the positive relationship between correlation and firing rate is related to the success of the low-rank approximation.

\section*{Discussion}

We simulated heterogeneous, asynchronous networks in order to investigate a possible relationship between firing rates and pairwise correlations in recurrent networks. We found that correlations can either 
increase \textit{or} decrease with firing rates; this could be attributed to differences in how cells responded to fluctuations in inhibitory conductances. When correlations \textit{did} increase with firing rates, this 
relationship was reflected in low-dimensional structure in the correlation matrix.

This study offers an example of a practical consequence of the difference between treating synaptic inputs as \textit{conductances} rather than \textit{currents}; while most synaptic currents are more accurately modeled as conductances, current-based formulations are often used for analytical and computational simplicity.
Although it is known that neural models responding to currents vs. conductances differ in their response dynamics \cite{lindner01,shelley02,at06}, 
this approach is supported by findings that steady-state firing rates are qualitatively similar in both settings (e.g. \cite{richardson07}). 
Here, we found that refined features of the steady-state firing rate surface will govern susceptibility to common input in asynchronous networks; two ``cuts" through this surface may yield divergent behavior with respect to correlation susceptibility, despite yielding similar firing rates.

In other words, the relationship between pairwise correlations and firing rate will depend on the means through which firing rate diversity is achieved.  In our study, we created firing rate diversity by regulating cell excitability; if we had instead varied mean inhibitory input (by varying the number of inhibitory connections) or a background excitatory current (which would model diversity in stimulus tuning from feed-forward inputs), we would likely have seen a different pattern.  Finally, the recurrent network will also shape the path the cells follow through the ``firing rate surface"; to generalize \cite{nature_Rocha_Doiron_ESB} to recurrent networks, we need to identify both how firing rates are produced and how they are shaped by the recurrent network. 

Low-dimensional structure has been a common finding in many large-scale neural recordings \cite{goris14,ecker14,lin15,kanashiro16,CY14}; while the origin is not always known, it is often interpreted as 
arising from a global input or top-down signal.
This is an interpretation that arises naturally from the technique of 
\textit{factor analysis}, in which one seeks to explain a data vector as the sum of a random vector and the linear combination of some number of latent factors \cite{GH1996} (for Gaussian random variables, 
each latent factor can literally be interpreted as a global input with a distinct pattern of projection onto the observed variables). 
In our network, we found that a single latent factor was effective at capturing correlations in the strong asynchronous regime; however, this latent factor did not reflect common input 
(there was no global external input into the network) but rather modulation from single-cell characteristics.  Thus, we identify a novel mechanism that can contribute to low-dimensional structure in neural recordings. 

\subsection*{Impact on stimulus coding}
The networks studied here were not encoding a stimulus; correlations were generated by recurrent activity, given that each neuron had a baseline firing rate in the absence of recurrent input.  However, we can readily connect this network to a stimulus coding task, in order to understand how the correlation-firing rate relationship can impact coding.  

Consider a population of cells that is responsible for encoding a single scalar stimulus $\theta$, such as movement direction or orientation of a visual stimulus, 
and that each cell has roughly a bell-shaped tuning curve. Furthermore, we model incoming stimulus by modulating a stimulus-dependent background current $I_{i,}(\theta)$; i.e., cells which prefer the current 
stimulus have a higher level of current, and thus a  higher firing rate, than cells which prefer an orthogonal or opposite stimulus. The network we studied here would model the response to a single stimulus $
\theta_0$; that is, the firing rate diversity we observe is present because some cells are strongly tuned to the current stimulus, while others are not.

We could extend this model, by resetting background currents to model a complete set of stimuli $\{\theta_1, \theta_2,...\theta_{n-1} \}$.  
For each stimulus $\theta_j$, correlations would show the rough firing rate dependence displayed in the strong asynchronous network, resulting in a \textit{stimulus-dependent} correlation structure in which pairwise correlations vary like geometric mean firing rate. This is the structure analyzed in \cite{zylberberg16,franke16}: the authors found that such a stimulus-dependent correlation code enhances information, when compared to a stimulus-independent code with the same average correlation level. Intuitively, the mean population response lives on the surface of a (hyper-)sphere in neural response space; the population encodes location on this surface.
Positive correlations between similarly tuned cells produce response distributions that are stretched in the radial direction, ``orthogonal" to this sphere, and thus have a minimal impact on the encoded variable.

Moreover, the mechanism that produced stimulus-dependent correlations in \cite{zylberberg16,franke16} was similar to that shown here (see also \cite{lin15}); common input modulated by stimulus-dependent gain factors. Here, we demonstrated how these stimulus-dependent gain factors might arise (or not) in a recurrent network. If excitation is tuned to put the network in the strong asynchronous regime, then the (stimulus-dependent) correlation structure that results will be favorable to coding. If excitation is tuned to put the network in the asynchronous regime, then correlations are overall low and not stimulus-dependent (although, given that average correlations are not matched, we do not here compare information contained within the two networks).

\subsection*{Future work}
This work has, necessarily, focused only on a subset of network attributes that might affect firing statistics. One important feature is the frequency of higher-order graph motifs; experiments have shown that 
specific motifs will occur more frequently, than would be expected 
in an Erd\H{o}s-R\'enyi network with fixed single-cell connection probability 
\cite{song05}.  
Theoretical work has found that in networks of integrate-and-fire neurons, an overabundance of divergent and chain motifs will lead to enhanced correlation \cite{Hu+2013} (this finding does depend on the dynamical regime; different motifs impact correlations in networks of coupled oscillators\cite{zhao_etal_FCN_2011}).
In \cite{Hu+2013}, the authors use the assumption of homogeneous single-cell characteristics to find parsimonious and instructive formulae for the average correlation, and give a roadmap for how this might be generalized to heterogeneous networks.  We look forward to considering the combined effect of single-cell \textit{and} network heterogeneity in future work.

Another source of cell-to-cell heterogeneity is how cells respond to stimuli, as emphasized in the previous discussion\cite{abbott99,moreno14,silveira14,zylberberg16,franke16} (see \cite{kohn16} for a review). Here, we did not consider a specific sensory system with tuning but rather focus on the general question of how the distribution of correlation values arise in recurrent networks. 
Given the previous discussion, one next step will be to investigate how correlations covary with firing rates, when cell-to-cell heterogeneity is produced by stimulus tuning in a structured network responding to a single variable (such as direction or orientation).
%


Finally, for numerical tractability our simulations here were performed in relatively small networks.
While high average correlations have been measured in experiments \cite{cohen11}, theoretical models of asynchronous networks have found that correlations must go to zero as the system becomes large ($N \rightarrow \infty$) \cite{renart10}. However, recent work has found that this does not have to be true, as long as spatial structure is introduced into the network \cite{rosenbaum_etal_2016}. 
We anticipate that this may carry over to other forms of heterogeneity, such as single-cell variability, and that therefore the effect we observe here persists for larger networks. We look forward to reporting on this in future work.


\section*{Methods}
\subsection*{Neuron model and network setup}\label{sec:meth1}
We considered randomly connected networks of excitatory and inhibitory neurons.
Each cell was a linear integrate-and-fire model with second-order alpha-conductances, i.e. membrane voltage $v_i$ was modeled with a stochastic differential equation, as long as it remained beneath a threshold $\theta_i$:
\begin{eqnarray}
\tau_m \frac{dv_i}{dt} &=& 	-v_i-  g_{E,i}(t)(v_i-\mathcal{E}_E) -  g_{I,i}(t)(v_i-\mathcal{E}_I) + \sigma_i \sqrt{\tau_m} \xi_i(t),   \label{eqn:LIF_diffeq}
\end{eqnarray}
When $v_i$ reaches $\theta_i$, it is reset to $0$ following a refractory period:
\begin{eqnarray} 
v_i(t + \tau_{\text{ref}}) &\rightarrow & 0, \qquad v_i(t)  \geq  \theta_i
\end{eqnarray}
Each neuron was driven by a Gaussian, white background noise, with magnitude $\sigma_i$ depending only on the cell type; that is, 
$\langle \xi_i(t) \rangle = 0$ and $\langle \xi_i(t) \xi_i (t+s) \rangle = \delta(s)$.
The membrane time constant, $\tau_m$, and excitatory and inhibitory synaptic reversal potentials, $\mathcal{E}_E$ and $\mathcal{E}_I$, are the same for every cell in the network.

Each cell responded to synaptic input through conductance terms, $g_{E,i}$ and $g_{I,i}$, which are each governed by a pair of differential equations:
\begin{eqnarray}
\tau_{d,X} \frac{dg_{X,i}}{dt} & = & -g_{X,i} + g^{(1)}_{X,i}\\
\tau_{r,X} \frac{dg^{(1)}_{X,i}}{dt} & = & -g^{(1)}_{X,i}  + \tau_{r,X} \alpha_{X} \left( \frac{W_{YX}}{N_{YX}} \right)\sum_{j\in X,j \rightarrow i} \sum_k\delta(t - t_{j,k})  \label{eqn:cond_aux_def}
\end{eqnarray}
where $Y = \{E,I\}$ denotes the type of cell $i$ and $X = \{E,I\}$ denotes the type of the source neuron $j$. Each spike is modeled as a delta-function that impacts the auxiliary variable $g^{(1)}_{X,i}$; here $t_{j,k}$ is the $k$-th spike of cell $j$. The rise and decay time constants $\tau_{r,X}$ and $\tau_{d,X}$ and pulse amplitude $\alpha_{X}$ depend only on the type of the source neuron; i.e. they are otherwise the same across the population.  The parameter $W_{YX}$ denotes the strength of $X \rightarrow Y$ synaptic connections, which are (once given the type of source and target neurons) identical across the population. 
The ``raw" synaptic weight (listed in Table~\ref{table:parameters}) is divided by $N_{YX}$, the total number of $X \rightarrow Y$ connections received by each $Y$-type cell. 

We chose connections to be homogeneous and relatively dense, consistent with the local architecture of cortex. 
Connection probabilities ranged from 20\,\%--40\,\%, consistent with experimentally measured values \cite{oswald09,yuste11,levy12}.
For our baseline network state, we then chose synaptic weights so the network is moderately inhibition-dominated ($\alpha_E W_{IE} <  \alpha_I W_{II}$ and $\alpha_E W_{EE} < \alpha_I W_{EI}$); that is both $E$ and $I$ cells receive more inhibition than excitation) and shows noisy spiking consistent with the classical asynchronous state. Each neuron receives a fixed number of incoming connections, the identities of which are chosen randomly. (The specific cell ID numbers differ in the different simulations shown below.) 
For most of the networks we discuss here $N=100$ with the 80/20 ratio typical of cortex (i.e $n_E = 80$, $n_I = 20$). Each excitatory cell receives $N_{EE} = 32$ (40 \%) excitatory and $N_{EI} = 7$ (35 \%) inhibitory connections; each inhibitory cell receives $N_{IE} = 16$ (20 \%) and $N_{II} = 8$ (40 \%) inhibitory projections.  

In heterogeneous networks, the threshold $\theta_i$ varied across the population. For both excitatory and inhibitory neurons, the thresholds $\theta_i$ were chosen from a log-normal distribution between $0.7$ and $1.4$ (where the rest potential, $V_r = 0$). To be precise,  $\log \theta_i$ was chosen from a (truncated) normal distribution with mean $-s^2_\theta/2$ and standard deviation $s_\theta$. With this choice, $\theta_i$ has mean 1 and variance: $e^{s_\theta^2}-1$. Thus we can view $s_\theta$ as a measure of the level of threshold heterogeneity.

Throughout this 
paper, we set $s_\theta=0.2$, which results in a wide range of firing rates compared to the homogeneous case.  
This was the only source of cell-to-cell heterogeneity; all other parameters were identical across the population, conditioned on neuron type. In homogeneous networks, the threshold was the same across the population: $\theta_i = 1$. 

\begin{table}
\centering
\caption{
{\bf Excitatory connection strengths mediate between different firing regimes.}}
\begin{tabular}{ | l | c | r | r | r | r | r | r | }
\hline
Parameter & $W_{EI}$ ($I \rightarrow E$)  & $W_{IE}$ ($E \rightarrow I$) & $W_{EE}$ & $W_{II}$ & $\sigma_i (i \in E)$ & $\sigma_i (i \in I)$\\
\hline
Asynchronous & 10 & 5 & 0.5 & 5 & $2/\sqrt{2}$ & $3/\sqrt{2}$ \\
Str. Asynch. & 10 & 8 & 9 & 5 & $1.5/\sqrt{2}$ &  $2.5/\sqrt{2}$  \\
\hline
\% connectivity & 35 \% & 20 \% & 40 \% & 40 \% & & \\
\hline
\end{tabular}
\begin{flushleft}
Here $W_{XY}$ denotes $X \rightarrow Y$ connections; e.g. $W_{EI}$ denotes the strength of excitatory connections onto inhibitory neurons. The parameter $\sigma_i$  denotes the strength of background noise in units of (scaled) voltage, and depends only on cell type ($E$ or $I$).
\end{flushleft}
\label{table:parameters}
\end{table} 

Monte Carlo simulations were performed using the stochastic forward- Euler method (Euler-Maruyama), with a time step much smaller than any time scale in the system ($\Delta t = 0.01$ ms). 
Each network was simulated for one second of simulation time, after an equilibration time. Then, a large number of realizations of this interval ($n_R =10^5$) were simulated. Spike counts were retained in each 1 ms window (for a total of 1000 windows) within a realization. 
\begin{table}
\centering
\caption{
{\bf Other parameters used in network simulations}}
\begin{tabular}{ | l | l | c | c |  }
\hline
Parameter & Definition & $X=E$ & $X=I$\\
\hline
$\tau_{r,X}$ & Synaptic rise time & 1 ms & 2 ms\\
$\tau_{d,X}$ & Synaptic decay time & 5 ms & 10 ms\\
$\tau_m$ & Membrane time constant & 20 ms & 20 ms\\
$\tau_{ref}$ & Refractory time & 2 ms & 2 ms\\
$\alpha_{X}$ & Pulse amplitude & 1 & 2\\
$\mathcal{E}_X$ & Synaptic reversal potential & 6.5 & -0.5\\
\hline
 \end{tabular}
\label{table:parameters2}
\end{table} 
With this large number of realizations/trials, the error bars on the resulting time-dependent firing rates were small. 
Therefore we emphasize that the firing rate pattern is largely driven by network connectivity; while firing is driven by random fluctuations in the background noise, any cell-to-cell variability in the \textit{trial-averaged} firing rates are not an artifact of the finite number of trials.



\subsection*{Linear Response Theory}  \label{sec:LR_theory_methods}
In general, computing the response of even a single neuron to an input requires solving a complicated, nonlinear stochastic process. However, it often happens that the presence of background noise linearizes the response of the neuron, so that we can describe this response as a perturbation from a background state. This response is furthermore linear in the perturbing input and thus referred to as \textit{linear response} theory \cite{risken}. The approach can be generalized to yield the dominant terms in the coupled network response, as well; we will use the theory to predict the covariance matrix of activity.


We first consider the case of a single cell: an LIF neuron responding to a mean zero current $\epsilon X_i (t)$
\[ \tau_m \frac{dv_i}{dt} = -(v_i-E_L) + E_i + \sigma_i \sqrt{\tau_m} \xi_i(t) +  \epsilon X_i(t).
\]
(otherwise, the mean of $X_i$ can simply be absorbed into $E_i$). 

For a fixed input $\epsilon X_i(t)$, the output spike train $y_i(t)$ will be slightly different for each realization of the noise $\xi_i(t)$ and initial condition $v_i(0)$. Therefore we try to work with the time-dependent firing \textit{rate}, $\nu_i(t) \equiv \langle y_i (t) \rangle$, which is obtained by averaging over all realizations and initial conditions.  Linear response theory proposes the ansatz that the firing rate can be described as a perturbation from a baseline rate proportional to the input $\epsilon X_i$:
\begin{eqnarray}
\nu_i (t) & = & \nu_{i,0} + (A_i * \epsilon X_i)(t); \label{eqn:susc_def}
\end{eqnarray}
$\nu_{i,0}$ is the baseline rate (when $X = 0$) and  $A_i(t)$ is a \textit{susceptibility function} that characterizes this firing rate response up to order $\epsilon$ 
\cite{lindner05,nature_Rocha_Doiron_ESB,trousdale12}. 

We now consider the theory for networks; here cell $i$ responds to the spike train of cell $j$, $y_j (t)$, via the synaptic weight matrix $\Wvec$, after convolution with a synaptic filter $F_j(t)$:
\[ \tau_m \frac{dv_i}{dt} = -(v_i-E_L) + E_i + \sigma_i \sqrt{\tau_m} \xi_i(t) +  \sum_j \Wvec_{ij} F_j * y_j(t) 
\]
In order to consider joint statistics, we need the trial-by-trial response of the cell. We first propose to approximate the response of each neuron as:
\begin{eqnarray} 
y_i(t) & \approx & y_i^0(t) + \left( A_i * \sum_j (\Jvec_{ij} * y_j ) \right) (t) ;
\end{eqnarray}
that is, each input $X_i$ has been replaced by the synaptic input,
and $\Jvec_{ij} = \Wvec_{ij} F_j(t)$ includes both the $i \leftarrow j$ synaptic weight $W_{ij}$ and synaptic kernel $F_j$ (normalized to have area 1); $A_i(t)$ is the susceptibility function from 
Eq~\ref{eqn:susc_def}.
In the frequency domain this becomes
\begin{eqnarray}
\tilde{y}_i(\omega) & = & \tilde{y}_i^0 + \tilde{A}_i(\omega) \left( \sum_j \tilde{\Jvec}_{ij}(\omega)  \tilde{y}_j (\omega) \right)
\end{eqnarray}
where $\tilde{y}_i = \mathcal{F}\left[ y_i - \nu_i \right]$ is the Fourier transform of the mean-shifted process ($\nu_i$ is the average firing rate of cell $i$) and $\tilde{f} = \mathcal{F}\left[ f \right]$ for all other quantities. In matrix form, this yields a self-consistent equation for $\tilde{y}$ in terms of $\tilde{y}^0$:
\begin{eqnarray}
\left( \Ivec - \tilde{\Kvec}(\omega) \right) \tilde{y}  =  \tilde{y}^0 & \Rightarrow & \tilde{y} = \left( \Ivec - \tilde{\Kvec}(\omega) \right)^{-1} \tilde{y}^0
\end{eqnarray}
where $\tilde{\Kvec}_{ij} (\omega) = \tilde{A}_i(\omega) \tilde{\Jvec}_{ij}(\omega)$ is the interaction matrix, in the frequency domain.
The cross-spectrum is then computed
\begin{eqnarray}
\langle \tilde{y}(\omega) \tilde{y}^{\ast}(\omega)  \rangle & = &  \left( \Ivec - \tilde{\Kvec}(\omega) \right)^{-1}  \langle \tilde{y}^0(\omega) \tilde{y}^{0 \ast}(\omega) \rangle  \left( \Ivec - \tilde{\Kvec}^{\ast}(\omega) \right)^{-1} \label{eqn:Xspec_LR}
\end{eqnarray}
To implement this calculation, we first solve for a self-consistent set of firing rates: that is, $\nu_i$ is the average firing rate of
\begin{eqnarray}
\tau_m \frac{dv_i}{dt}&  = & -(v_i-E_L) + (E_i + \Ex[f_i]) + \sigma_i \sqrt{\tau_m} \xi_i(t)    \label{eqn:self_consistent_r}
\end{eqnarray}
where $\Ex[f_i] = \sum_j \Wvec_{ij} \nu_j$.

We must then compute the power spectrum $\langle \tilde{y}^0(\omega) \tilde{y}^{0 \ast}(\omega) \rangle$ and the susceptibility $A_i(\omega)$, which is the (first order in $\epsilon$) response in the firing rate $r_i(t) = r_i^0 + \epsilon A_i(\omega) \exp( \imath \omega t)$ in response to an input current perturbation $X(t) = \epsilon \exp(\imath \omega t)$ (here $\imath$ is used for $\sqrt{-1}$, while $i$ denotes an index). Both can be expressed as the solution to (different) first-order boundary value problems and solved via Richardson's threshold integration method \cite{richardson07,richardson08}.

In our simulations, we used conductance-based neurons; 
this requires modification, compared with the simpler current-based models.  
We first approximate each conductance-based neuron as an effective current-based neuron with reduced time constant, following the discussion in \cite{Ger+2014}. 
First, separate each conductance into mean and fluctuating parts;  e.g.  $g_{E,i} \rightarrow \langle g_{E,i} \rangle + \left( g_{E,i} -  \langle g_{E,i} \rangle \right)$. 
%
Then we identify an effective conductance $g_{0,i}$ and potential $\mu_{i}$, and treat the fluctuating part of the conductances as noise, i.e. $g_{E,i} -  \langle g_{E,i} \rangle \rightarrow \sigma_{E,i} \xi_{E,i}(t)$:
\begin{eqnarray}
\tau_m \frac{dv_i}{dt} & = & - g_{0,i} (v_i - \mu_i) + \sigma_{E,i} \xi_{E,i}(t) (v_i - \mathcal{E}_E) + \sigma_{I,i} \xi_{I,i}(t)  (v_i - \mathcal{E}_I) + \sqrt{\sigma_i^2 \tau_m} \xi_i(t)  \label{eqn:eff_before_g0}
\end{eqnarray}
where
\begin{eqnarray}
g_{0,i} & = & 1 + \langle g_{E,i}\rangle + \langle g_{I,i} \rangle \\
\mu_i & = &  \frac{E_{L}+E_i + \langle g_{E,i} \rangle \mathcal{E}_E + \langle g_{I,i} \rangle \mathcal{E}_I}{g_{0i}}\\
\sigma_{E,i}^2 & = & \Var \left[ g_{E,i}(t) \right] = \Ex \left[ \left( g_{E,i}(t) - \langle g_{E,i} \rangle \right) ^2 \right]\\
\sigma_{I,i}^2 & = & \Var \left[ g_{I,i}(t) \right] = \Ex \left[ \left( g_{I,i}(t) - \langle g_{I,i} \rangle \right) ^2 \right]
\end{eqnarray}
We next simplify the noise terms by writing
\begin{eqnarray}
v_i - \mathcal{E}_E & = & v_i - \mu_i + \mu_i - \mathcal{E}_E
\end{eqnarray}
and assume that the fluctuating part of the voltage, $v_i - \mu_i$, is mean-zero and uncorrelated with the noise terms $\xi_{E,i}(t)$ \cite{Ger+2014}. That allows us to define an effective equation
\begin{eqnarray}
\tau_{{\rm eff},i} \frac{dv_i}{dt} & = &  - (v_i - \mu_i) + \sqrt{\sigma_{{\rm eff},i}^2 \tau_{{\rm eff},i}} \eta_{{\rm eff},i}(t)  \label{eqn:eff_for_cond_based}
\end{eqnarray}
where
\begin{eqnarray}
\tau_{{\rm eff},i} & = & \frac{\tau_m}{g_{0,i}}\\
\sigma_{{\rm eff},i}^2 & = & \frac{\sigma_{E,i}^2 (\mu_i - \mathcal{E}_E)^2 + \sigma_{I,i}^2 (\mu_i - \mathcal{E}_I)^2 + \sigma_i^2 \tau_m }{g_{0,i} \tau_m} 
\end{eqnarray}
and the fluctuating voltage, $v_i(t)-\mu_i$, now makes no contribution to the effective noise variance.

Finally, we consider how to model the conductance mean and variance, e.g.  $\langle g_{E,i} \rangle$ and $\sigma_{E,i}^2$.
In our simulations, we used second order $\alpha$-functions: each conductance $g_{X,i}$ is modeled by two equations that take the form
\begin{eqnarray}
\tau_{r,X} \frac{dg^{(1)}_{X,i}}{dt} & = & -g^{(1)}_{X,i} + \tau_{r,X} \hat{\alpha}_{X,i} \sum_{k} \delta(t-t_k)\\
\tau_{d,X} \frac{dg_{X,i}}{dt} & = & -g_{X,i} + g^{(1)}_{X,i}  \label{eqn:G}
\end{eqnarray}
where $X = E,I$ and the summation is over all type-$X$ spikes incoming to cell $i$. 
(For notation purposes, $\hat{\alpha}_{X,i}$ includes all factors that contribute to the pulse size in Eq~\ref{eqn:cond_aux_def}, including synapse strength and pulse amplitude.) 
The time constants $\tau_{r,X}$, $\tau_{d,X}$ may depend on synapse type; the spike jumps $\hat{\alpha}_{X,i}$ may depend on synapse type and target cell identity.
We assume that each spike train is Poisson, with a constant firing rate: i.e. each spike train is modeled as a stochastic process $S(t)$ with
\begin{eqnarray*}
\langle S(t) \rangle & = & \nu\\
\langle S(t)S(t+\tau) \rangle - \nu^2 & = & \nu \delta(\tau)
\end{eqnarray*}
Then a straightforward but lengthy calculation shows that
\begin{eqnarray}
\langle g_{X,i}(t) \rangle & = &  \hat{\alpha}_{X,i} \nu_{X,i} \tau_{r,X}   \label{eqn:Gmean}\\
\Var \left[ g_{X,i}(t) \right] & = &  \left( \frac{1}{2} \hat{\alpha}_{X,i}^2 \nu_{X,i} \tau_{r,X}  \right) \left( \frac{\tau_{r,X}}{\tau_{r,X} + \tau_{d,X}} \right) \label{eqn:mean_var_G_2nd_order_alpha}
\end{eqnarray}
where $\nu_{X,i}$ is the total rate of type-$X$ spikes incoming to cell $i$.

We now describe how these considerations modify the linear response calculation. First, for the self-consistent firing rate calculation, Eq~\ref{eqn:self_consistent_r} is replaced by an equation with a modified time constant, conductance, and noise (Eq~\ref{eqn:eff_for_cond_based}).

We next compute the susceptibility in response to parameters associated with the conductance, i.e. $\langle g_{E,i} \rangle$ and $\sigma_{E,i}^2$.  This differs from the current-based case in two ways: first, there is voltage-dependence in the diffusion terms, which results in a different Fokker-Planck equation (and thus a different boundary value problem to be solved for the power spectrum $\langle \tilde{y}^0(\omega) \tilde{y}^{0 \ast}(\omega) \rangle$). Second, modulating the rate of an incoming spike train will impact \textit{both} the mean and variance of the input to the effective equation, 
Eq~\ref{eqn:eff_before_g0} (via $\mu_i$ and $\sigma_{X,i}$). 
Furthermore, this impact may differ for excitatory and inhibitory neurons, giving us a total of \textit{four} parameters that can be varied in the effective equation.
However, neither consideration presents any essential difficulty \cite{richardson07}.  

Therefore we apply Richardson's threshold integration method directly to Eq~\ref{eqn:eff_before_g0}:
\begin{eqnarray}
\tau_m \frac{dv_i}{dt} & = & - g_{0,i} (v_i - \mu_i) + \sigma_{E,i} \xi_{E,i}(t) (v_i - \mathcal{E}_E) + \sigma_{I,i} \xi_{I,i}(t)  (v_i - \mathcal{E}_I) + \sqrt{\sigma_i^2 \tau_m} \xi_i(t)  \label{eqn:eff_before_g0_v2}
\end{eqnarray}
When we compute susceptibilities, the parameter to be varied is either a mean conductance --- $\langle g_{E,i} \rangle \rightarrow \langle g_{E,i} \rangle_0 + \langle g_{E,i} \rangle_1 \exp(\imath \omega t)$ or $\langle g_{I,i} \rangle \rightarrow \langle g_{I,i} \rangle_0 + \langle g_{I,i} \rangle_1 \exp(\imath \omega t)$ --- or a variance --- $\sigma_{E,i}^2 \rightarrow \left( \sigma_{E,i}^2 \right)_0 +  \left( \sigma_{E,i}^2 \right)_1 \exp(\imath \omega t)$ or  $\sigma_{I,i}^2 \rightarrow \left( \sigma_{I,i}^2 \right)_0 +  \left( \sigma_{I,i}^2 \right)_1 \exp(\imath \omega t)$.
Thus we have a total of four susceptibility functions $\tilde{A}_{\langle g_E \rangle, i}(\omega)$, $\tilde{A}_{\langle g_I \rangle, i}(\omega)$, $\tilde{A}_{\sigma_E^2, i}(\omega)$, and $\tilde{A}_{\sigma_I^2, i}(\omega)$. Since the Fokker-Planck equation to be solved is linear, we can compute both susceptibilities separately and then add their effects. We now have the interaction matrix:
\begin{eqnarray}
\tilde{\Kvec}_{ij} (\omega) & = & \left\{ \begin{array}{ll} 
	\tilde{A}_{\langle g_E \rangle, i}(\omega) \tilde{\Jvec}_{ij}(\omega) + \tilde{A}_{\sigma_E^2, i}(\omega) \tilde{\Lvec}_{ij}(\omega), & \qquad j \, {\rm excitatory} \\
	\tilde{A}_{\langle g_I \rangle, i}(\omega) \tilde{\Jvec}_{ij}(\omega) + \tilde{A}_{\sigma_I^2, i}(\omega) \tilde{\Lvec}_{ij}(\omega), & \qquad j \, {\rm inhibitory}
	\end{array} \right.
\end{eqnarray}
where $\tilde{\Lvec}(\omega)$ plays a similar role as $\tilde{\Jvec}$, but for the effect of incoming spikes on the \textit{variance} of conductance. Its relationship to $\tilde{\Jvec}$ (either in the frequency or time domain) is given by the same simple scaling shown in Eq~\ref{eqn:mean_var_G_2nd_order_alpha}: i.e., for $j$ excitatory, 
\begin{eqnarray}
\tilde{\Lvec}_{ij} (\omega) & = & \tilde{\Jvec}_{ij}(\omega) \times  \left( \frac{\hat{\alpha}_{E,i}}{2} \right) \times \left( \frac{\tau_{r,E}}{\tau_{r,E} + \tau_{d,E}} \right)  \label{eqn:L_ij}
\end{eqnarray}
where the first factor comes from the effective spike amplitude $\hat{\alpha}_{E,i}$ (and is the scale factor proposed in \cite{richardson07}, Eqn. (64)), and the second arises from using second-order (vs. first-order) alpha-functions.

We use a modified version of the implementation given by \cite{trousdale12} for Richardson's threshold integration algorithm \cite{richardson07,richardson08} to compute rate $\nu_i$, power $\langle \tilde{y}_i^0(\omega) \tilde{y}_i^{0 \ast}(\omega) \rangle$, and the various susceptibilities ($\tilde{A}_{\langle g_E \rangle, i}(\omega)$, $\tilde{A}_{\langle g_I \rangle, i}(\omega)$, $\tilde{A}_{\sigma_E^2, i}(\omega)$, and $\tilde{A}_{\sigma_I^2, i}(\omega)$) for an LIF neuron. We validated our code using exact formulas known for the LIF 
\cite{ricciardi77}, and qualitative results from the literature \cite{lindner}.  

\subsection*{Computing statistics from linear response theory} \label{sec:stats_def}
 Linear response theory yields the cross spectrum of the spike train, $\langle \tilde{y}_i(\omega) \tilde{y}_j^{\ast}(\omega)  \rangle$, for each distinct pair of neurons $i$ and $j$ (see 
 Eq~\ref{eqn:Xspec_LR}). To recover a representative set of statistics, 
 we rely on several standard formulae relating this function to other statistical quantities.

The cross correlation function, $\Cvec_{ij}(\tau)$, measures the similarity between two processes at time lag $\tau$, while the cross spectrum measures the similarity between two processes at frequency $\omega$:
 \begin{eqnarray}
 \Cvec_{ij} (\tau) & \equiv & \langle (y_i(t)-\nu_i)(y_j(t + \tau)-\nu_j) \rangle \\
 \tilde{\Cvec}_{ij}(\omega) & \equiv & \langle \tilde{y}_i (\omega) \tilde{y}_j (\omega) \rangle   \label{eqn:Xspec_def}
 \end{eqnarray}
The Weiner-Khinchin theorem \cite{risken} implies that $\{ \Cvec_{ij}, \tilde{\Cvec}_{ij} \} $ are a Fourier transform pair: that is,
 \begin{eqnarray}
 \tilde{\Cvec}_{ij} (\omega) & = & \int_{-\infty}^{\infty} \Cvec_{ij} (t) e^{-2 \pi \imath \omega t} \, dt  
 \end{eqnarray}
 
 In principle, the crosscorrelation $\Cvec(t)$ and cross-spectrum  $\tilde{\Cvec}(\omega)$ matrices are functions on the real line, reflecting the fact that correlation can be measured at different time scales.  In particular, for a 
 stationary point process the covariance of spike counts over a window of length $T$, $n_i$ and $n_j$, can be related to the crosscorrelation function $\Cvec_{ij}$ by the following formula \cite{kaybook}:
\begin{eqnarray}
\Cov_T(n_i, n_j) & = &\int_{-T}^{T} \Cvec_{ij}(\tau) \left(T - \mid \tau \mid \right) \, d\tau
\end{eqnarray}
The variance of spike counts over a time window of length $T$, $n_i$, is likewise given by integrating the autocorrelation function $\Cvec_{ii}$:
\begin{eqnarray}
\Var_T(n_i) & = &\int_{-T}^{T} \Cvec_{ii}(\tau) \left(T - \mid \tau \mid \right) \, d\tau
\end{eqnarray}

It can be helpful to  normalize by the time window, i.e.
\begin{eqnarray}
\frac{\Cov_T(n_i, n_j)}{T} & = & \int_{-T}^{T} \Cvec_{ij}(\tau) \left(1 - \frac{\mid \tau \mid}{T} \right) \, d\tau;
\end{eqnarray}
we can now see that for an integrable cross correlation function (and bearing in mind that the cross-spectrum is the Fourier transform of the cross correlation), that
\begin{eqnarray}
\lim_{T \rightarrow \infty} \frac{\Cov_T(n_i, n_j)}{T} & = & \int_{-\infty}^{\infty} \Cvec_{ij}(\tau) d\tau \; = \; \tilde{\Cvec}_{ij}(0)  \label{Xspec_longT}
\end{eqnarray}
while
\begin{eqnarray}
\lim_{T \rightarrow 0} \frac{\Cov_T(n_i, n_j)}{T^2} & = &  \frac{1}{T} \int_{-T}^{T} \Cvec_{ij}(\tau) \left(1 - \frac{\mid \tau \mid}{T} \right) \, d\tau   \; \approx  \; \Cvec_{ij}(0) 
\end{eqnarray}
Thus, we can use $\tilde{\Cvec}_{ij}(0)$ and $\Cvec_{ij}(0)$ as measures of long and short time correlations respectively.

Finally, the Pearson's correlation coefficient of the spike count defined as:
\begin{eqnarray}
\rho_{T,ij} & = & \frac{\Cov_T(n_i,n_j)}{\sqrt{\Var_T(n_i) \Var_T(n_j)}}  \label{eqn:rho_def}
\end{eqnarray}
is a common normalized measure of noise correlation, with $\rho \in [-1, 1]$. While $\Cov_T$ and $\Var_T$ grow linearly with $T$ (for a Poisson process, for example), $\rho_{T,ij}$ in general will not (although it may increase with $T$). 
In general, $\rho_{T,ij}$ depends on the time window $T$; however for readability we will often suppress the $T$-dependence in the notation (and use $\rho_{ij}$ instead).

\subsection*{Quantifying the role of motifs in networks} \label{sec:motif_def}
We next explain how we can use the results of linear response theory to give insight into the role of different paths in the network.
We begin with our predicted cross-spectrum (Eqs~\ref{eqn:Xspec_LR}, \ref{eqn:Xspec_def}) and apply a standard series expansion for the matrix inverse:
\begin{eqnarray}
\tilde{\Cvec}(\omega) & = &  \left( \Ivec - \tilde{\Kvec}(\omega) \right)^{-1}  \tilde{\Cvec}^0(\omega)   \left( \Ivec - \tilde{\Kvec}^{\ast}(\omega) \right)^{-1}  \\    \label{eqn:Xspec_LR_v2}
& = & \left[ \sum_{k=0}^{\infty} \left( \tilde{\Kvec}(\omega) \right)^k \right] \tilde{\Cvec}^0(\omega) \left[ \sum_{l=0}^{\infty} \left( \tilde{\Kvec}(\omega) \right)^l \right] \\
& = & \sum_{k=0}^{\infty} \sum_{l=0}^{\infty} \left( \tilde{\Kvec}(\omega) \right)^k \tilde{\Cvec}^0(\omega) \left( \tilde{\Kvec}(\omega)  \right)^l  \label{eqn:Cinf_def}
\end{eqnarray}
where $\tilde{\Cvec}^0(\omega)$ is a diagonal matrix containing the power spectra of the unperturbed processes; i.e. $\tilde{\Cvec}^{0}_{ii} \equiv \langle  \tilde{y}_i (\omega) \tilde{y}_i (\omega) \rangle $.
This double sum will converge as long as the spectral radius of $\tilde{\Kvec}$ is less than 1 \cite{trousdale12}. 

By truncating this double sum to contain terms such that $k + l \le n$, we define the $n$th approximation to the cross-spectrum:
\begin{eqnarray}
\tilde{\Cvec} (\omega) & \approx & \tilde{\Cvec}^n (\omega) \\
& = & \tilde{\Cvec}^0(\omega) + \sum_{k=1}^n \left[ \sum_{l=0}^k \left( \tilde{\Kvec}(\omega)\right)^{k-l} \tilde{\Cvec}^0(\omega) \left(\tilde{\Kvec}^*(\omega)\right)^{l}  \right] \label{eqn:Cn_def}
\end{eqnarray}
Each distinct term in the inner sum can be attributed to a particular undirected path of length $k$. Terms of the form $\tilde{\Kvec}^{k} \tilde{\Cvec}^0$ and $\tilde{\Cvec}^0 \left( \tilde{\Kvec}^*\right)^{k} $ account for unidirectional paths from $j\rightarrow i$ and $i \rightarrow j$ respectively;  the term $\left( \tilde{\Kvec}(\omega)\right)^{k-l} \tilde{\Cvec}^0(\omega) \left(\tilde{\Kvec}^*(\omega)\right)^{l}$ captures the contribution from a cell that has a length $l$ path onto cell $j$ and a length $k-l$ path onto cell $i$. Thus, we can use Eq~\ref{eqn:Cn_def} to decompose the correlation into contributions from different motifs (\cite{pernice11}, see also \cite{ostojic09}, \cite{rangan_PRL_2009}). 

We can also consider the 
contribution from all length-$n$ paths; that is, 
\[
\tilde{\Pvec}^n = \tilde{\Cvec}^n (\omega) - \tilde{\Cvec}^{n-1} (\omega)  = \sum_{l=0}^n \left( \tilde{\Kvec}(\omega)\right)^{n-l} \tilde{\Cvec}^0(\omega) \left(\tilde{\Kvec}^*(\omega)\right)^{l}
\]
If the sum in Eq~\ref{eqn:Cinf_def} converges, 
we should expect the magnitude of contributions to decrease as $n$ increases. 

We will also show the \textit{normalized}
contribution from length-$n$ paths, 
which we define as follows: let $\Lamvec (\omega)$ be the diagonal matrix with $\Lamvec_{ii}(\omega) = \tilde{\Cvec}_{ii}(\omega)$. Then we define the matrix of contributions from length-$n$ paths $\tilde{\Rvec}^n$ as follows:
\begin{eqnarray}
\tilde{\Rvec}^n (\omega) & = & \Lamvec^{-1/2}(\omega) \tilde{\Pvec}^n(\omega)  \Lamvec^{-1/2}(\omega) 
\end{eqnarray}
Equivalently, $\tilde{\Rvec}^n_{ij}(\omega) = \tilde{\Pvec}^{n}_{ij}(\omega)/\sqrt{\tilde{\Cvec}_{ii}(\omega) \tilde{\Cvec}_{jj}(\omega)}$.
This effectively normalizes the cross correlation by the autocorrelation; in particular, we can use this to decompose the correlation coefficient (Eq~\ref{eqn:rho_def}) for long time windows, because
$\lim_{n\rightarrow \infty} \sum_{k=0}^n \tilde{\Rvec}^k(0) = \lim_{T \rightarrow \infty} \rho_{T,ij}$. 

In general, we will show long-timescale correlation (e.g.  $\tilde{\Cvec}(0)$ or $\tilde{\Rvec}^n(0)$) (Eq~\ref{Xspec_longT}); 
results were qualitatively similar for other timescales.

\subsection*{Quantifying correlation susceptibility} \label{sec:corr_susc}
We next consider how to quantify the (linear) susceptibility of correlation to a change in parameter. Returning to Eq~\ref{eqn:susc_def}, but written in terms of the single-cell response:
\begin{eqnarray}
y_i (t) & = & y_{i,0} + (A_{\mu,i} * X_{\mu})(t)  \qquad \Rightarrow\\  \label{eqn:susc_def_y}
\tilde{y}_i (\omega) & = & \tilde{y}_{i,0}(\omega) + \tilde{A}_{\mu,i} (\omega) \tilde{X}_{\mu} (\omega)  
\end{eqnarray}
Here, $X_{\mu}(t)$ is a (possibly) time-dependent change in a parameter, such as input current or mean inhibitory conductance; $y_{i,0}$ is the baseline spike train (when $X = 0$).  $A_{\mu,i}(t)$ is a \textit{susceptibility function} that characterizes the cell's response (to the parameter variation) as long as $X_{\mu}(t)$ is small
\cite{lindner05,nature_Rocha_Doiron_ESB,trousdale12}.  Following \cite{nature_Rocha_Doiron_ESB}, the cross-spectrum of $y$ can now be approximated as:
\begin{eqnarray}
\tilde{\Cvec}_{ij}(\omega) & \equiv & \langle  \tilde{y}^{*}_{i} \tilde{y}_{j} \rangle \approx \langle  \tilde{y}^{*}_{i,0} \tilde{y}_{j,0} \rangle +  \langle  \tilde{A}^{*}_{\mu,i} \tilde{X}^*_{\mu} \tilde{y}_{j,0} \rangle +  \langle  \tilde{A}_{\mu,j} \tilde{X}_{\mu} \tilde{y}^*_{i,0} \rangle +  \tilde{A}^{*}_{\mu,i}  \tilde{A}_{\mu,j}  \langle \tilde{X}^{*}_{\mu} \tilde{X}_{\mu} \rangle \\
& = & \tilde{A}^{*}_{\mu,i} (\omega) \tilde{A}_{\mu,j}(\omega) \tilde{C}_{\mu}(\omega) 
\end{eqnarray}
where $\tilde{C}_{\mu}(\omega)$ is the spectrum of the parameter variation.
The susceptibility has an appealing interpretation in the limit $\omega \rightarrow 0$, as the derivative of the classical \textit{f-I curve}:
\begin{eqnarray}
\lim_{\omega \rightarrow 0} \tilde{A}_{\mu,i}(\omega) & = & \frac{d \nu_i}{d \mu}
\end{eqnarray}
where $\nu_i$ is the steady-state firing rate of cell $i$, assuming we can measure it for specific values of the parameter $\mu$.
\begin{eqnarray}
\lim_{T \rightarrow \infty} \rho_{T,ij} & = & \lim_{T \rightarrow \infty} \frac{\Cov_T(n_i,n_j)}{\sqrt{\Var_T(n_i) \Var_T(n_j)}}  = \frac{\tilde{\Cvec}_{ij}(0)}{\sqrt{\tilde{\Cvec}_{ii}(0) \tilde{\Cvec}_{jj}(0)}}\\
& \approx & \frac{\tilde{A}_{\mu,i}(0) \tilde{A}_{\mu,j}(0)}{\sqrt{\tilde{\Cvec}_{ii}(0) \tilde{\Cvec}_{jj}(0)}} \tilde{C}_{\mu}(0) 
\end{eqnarray}
This motivates the definition of a \textit{correlation susceptibility}, which approximates the change in pairwise correlation induced by a parameter change experienced by both cells $i$ and $j$: 
\begin{eqnarray}
S^{\mu}_{ij} & = & \frac{\tilde{A}_{\mu, i}(0) \tilde{A}_{\mu, j}(0)}{\sqrt{\tilde{\Cvec}_{ii}(0)\tilde{\Cvec}_{jj}(0)}}
\end{eqnarray}
If this increases with firing rate --- that is, if $\frac{d S^{\mu}_{ij} }{d \nu} > 0$ --- then corrections will also increase with firing rate.

We can further analyze this quantity by making an assumption for asynchronous spiking, that spike count variance is equal to spike count mean; i.e. $\Var_T(n_i) = T \nu_i \Rightarrow \tilde{\Cvec}_{ii} = \nu_i$. Then
\begin{eqnarray}
S^{\mu}_{ij} & \approx & \frac{1}{\sqrt{\nu_i \nu_j}} \tilde{A}_{\mu, i}(0) \tilde{A}_{\mu, j}(0) = \frac{\tilde{A}_{\mu, i}(0)}{\sqrt{\nu_i }}  \frac{\tilde{A}_{\mu, j}(0)}{\sqrt{\nu_j }} 
\end{eqnarray}
which motivates the definition of the single-cell quantity
\[ S^{\langle g_I \rangle}_{i} \equiv \frac{\tilde{A}_{\langle g_I \rangle,i}(0)}{\sqrt{\nu_i}}  \]
In general, the firing rate depends on \textit{all} single cell parameters included in Eqn. ; i.e. there exists some function $f$ such that
\begin{eqnarray} 
\nu_i & = & f \left( \langle g_{I,i} \rangle, \sigma_{I,i}, \langle g_{E,i} \rangle, \sigma_{E,i}, \sigma_i, \theta_i \right) \\
\tilde{A}_{\langle g_I \rangle,i}(0) & = & \frac{\partial f}{\partial x_1}\left( \langle g_{I,i} \rangle, \sigma_{I,i}, \langle g_{E,i} \rangle, \sigma_{E,i}, \sigma_i, \theta_i \right) 
\end{eqnarray}
(recall that the susceptibility for $\omega = 0$ is the derivative of the firing rate with respect to the appropriate parameter (here, mean inhibitory conductance $\langle g_I \rangle$).

\subsection*{Low-rank approximation to the correlation matrix}  \label{sec:lowrank_def}
We consider the correlation matrix of spike counts, 
as measured from Monte Carlo simulations; while these are in principle related to the cross-correlation functions $\Cvec(t)$ defined in \textbf{Methods: Computing statistics from linear response theory}
we will 
use $\Cvec_T$ to denote the matrix of correlation coefficients measured for time window $T$; i.e.
\begin{eqnarray}
\left(\Cvec_T \right)_{ij} & = & \rho_{T, ij}
\end{eqnarray}
Furthermore, we will restrict to the E-E correlations; i.e. $\Cvec_T$ will be a $n_E \times n_E$ matrix, with ones on the diagonal (as $\rho_{T,ii} = 1$).

When we examined the singular values of the E-E correlation matrices obtained from Monte Carlo simulations, we noticed a consistent trend: there was usually one large cluster with one positive outlier. This motivates the following simple idea: by subtracting off a multiple of the identity matrix, $\lambda \Ivec$, we shift the cluster towards zero;  consequently $\Cvec_T - \lambda \Ivec$ is close to a rank-1 matrix.   We then propose to use the sum of the two 
as an approximation to $\Cvec_T$:
\begin{equation}
	\Cvec_T \approx \lambda \Ivec + (\sigma_1-\lambda) \uvec_1 \uvec_1^T.
\end{equation}

We seek the value $\lambda$ which maximizes the fraction of the Frobenius norm explained by the first singular vector: i.e. in terms of the singular values, 
\begin{eqnarray}
\lambda & = & \max_{\lambda} \frac{\tilde{\sigma}_1^2}{\displaystyle\sum_{j=1}^r \tilde{\sigma}_j^2}\\
& = & \max_{\lambda} \frac{(\sigma_1-\lambda)^2}{\displaystyle\sum_{j=1}^r (\sigma_j-\lambda)^2}
\end{eqnarray}
Since $\Cvec_T$ is symmetric semi-positive definite, the singular values $\sigma_j$ are equal to the eigenvalues $\lambda_j$: here $\sigma_1 \ge \sigma_2 \ge \cdots \ge \sigma_r \ge 0$ and $r$ is the rank of $\Cvec_T$. This has an exact solution: 
\begin{eqnarray}
\lambda & = & \lambda_1 - \frac{\displaystyle\sum_{j>1} (\lambda_1 - \lambda_j)^2}{\displaystyle\sum_{j>1} \lambda_1 - \lambda_j}
\end{eqnarray}
Because we have subtracted a multiple of the identity matrix, none of the singular vectors will have changed. 
We then have 
\begin{eqnarray}
\Cvec_T & \equiv & \lambda \Ivec + (\Cvec_T - \lambda \Ivec)\\
& = & \lambda \Ivec + \sum_{i=1}^r (\sigma_i - \lambda) \uvec_i \uvec_i^T
\end{eqnarray}
By truncating this sum, we approximate $C$ with a shifted low-rank matrix:
\begin{eqnarray}
\Cvec_T  \approx \Cvec_T^{\rm diag + R1} & \equiv & \lambda \Ivec + (\sigma_1 - \lambda) \uvec_1 \uvec_1^T
\end{eqnarray}
This procedure is similar to \textit{factor analysis}, in which one seeks to explain a data vector as the sum of a random vector ($\uvec$) and the linear combination of some number of latent factors ($\zvec$) \cite{GH1996}:
\begin{eqnarray*}
\xvec & = & \Lambda \zvec + \uvec;
\end{eqnarray*} 
the entries of $\xvec$ would then have the correlation matrix $\Psi + \Lambda \Lambda^T$, where $\Psi$ is a diagonal matrix containing the variances of $\uvec$.

\section*{Supporting Information}
%
%

\paragraph*{S1 Text:} Includes supplementary analysis of statistics and numerical methods, including discussion of supplementary figures.

\paragraph*{S1 Fig. Theory predicts population statistics in the asynchronous regime.}
\paragraph*{S2 Fig. Theory predicts population statistics in the strong asynchronous regime.}
\paragraph*{S3 Fig. Theory predicts cell-by-cell statistics in the asynchronous regime.}
\paragraph*{S4 Fig. Theory predicts cell-by-cell statistics in the strong asynchronous regime.}
\paragraph*{S5 Fig. Theory captures low-rank structure in correlation matrices.}
\paragraph*{S6 Fig. Effective parameters in the heterogeneous network: asynchronous regime} 
\paragraph*{S7 Fig. Effective parameters in the heterogeneous network: strong asynchronous regime} 
\paragraph*{S8 Fig. Firing rate as a function of inhibitory conductance and threshold.} 

\paragraph*{S1 Table. Statistics from heterogeneous vs. homogeneous networks: asynchronous regime}
\paragraph*{S2 Table. Statistics in recurrent networks: Monte Carlo vs. linear response theory, asynchronous regime}
\paragraph*{S3 Table. Statistics from heterogeneous vs. homogeneous networks: strong asynchronous regime}
\paragraph*{S4 Table. Statistics in recurrent networks: Monte Carlo vs. linear response theory, strong asynchronous regime}

\section*{Acknowledgments}
We thank Kre\v{s}imir Josi\'{c} for helpful comments on an earlier version of this manuscript, and Brent Doiron and Eric Shea-Brown for helpful conversations.


\nolinenumbers

%
%
%

\bibliography{Corr_vs_FR_R1_forArxiv}

\clearpage

\renewcommand\thefigure{S\arabic{figure}}    
\setcounter{figure}{0}   
 
\setcounter{table}{0}
\renewcommand{\thetable}{S\arabic{table}}

\noindent
{\Large Supplementary Information for \\
``When do Correlations Increase with Firing Rates?"}\\

\noindent
Andrea K. Barreiro and Cheng Ly

\subsection*{Linear response theory predicts the distribution of first- and second-order statistics in recurrent networks}\label{S2_Appendix}

\noindent
In recurrent networks, the response of each cell is shaped by both direct and indirect connections through the network. To separate the impact of different network mechanisms, we applied a network linear response theory (described in {\bf Methods: Linear Response Theory})
which allows us to decompose network correlations into contributions from different graph motifs (as in \cite{trousdale12,pernice11}). 
Here, we verify that this theory accurately predicted the results of Monte Carlo simulations.  

The network connectivity matrix $\Wvec$ and all other parameter values were the same as used in 
Monte Carlo simulations;
linear response theory yields a predicted value for the stationary firing rate $\nu_i$ and spike count variance $\Var_T[n_i]$ of each cell $i$, as well as the spike count covariance of each distinct cell pair, $\Cov_T(n_i, n_j)$. 
For each distinct network, we then compared the population distribution of single cell firing rates $\nu_i$, spike count variances, and two-cell covariances (as well as two-cell correlation coefficients $\rho^{EE}$ and $\rho^{IE}$), with the population distributions we obtained from Monte Carlo simulations.
\begin{figure}[h]
\centering
\includegraphics[width=\textwidth]{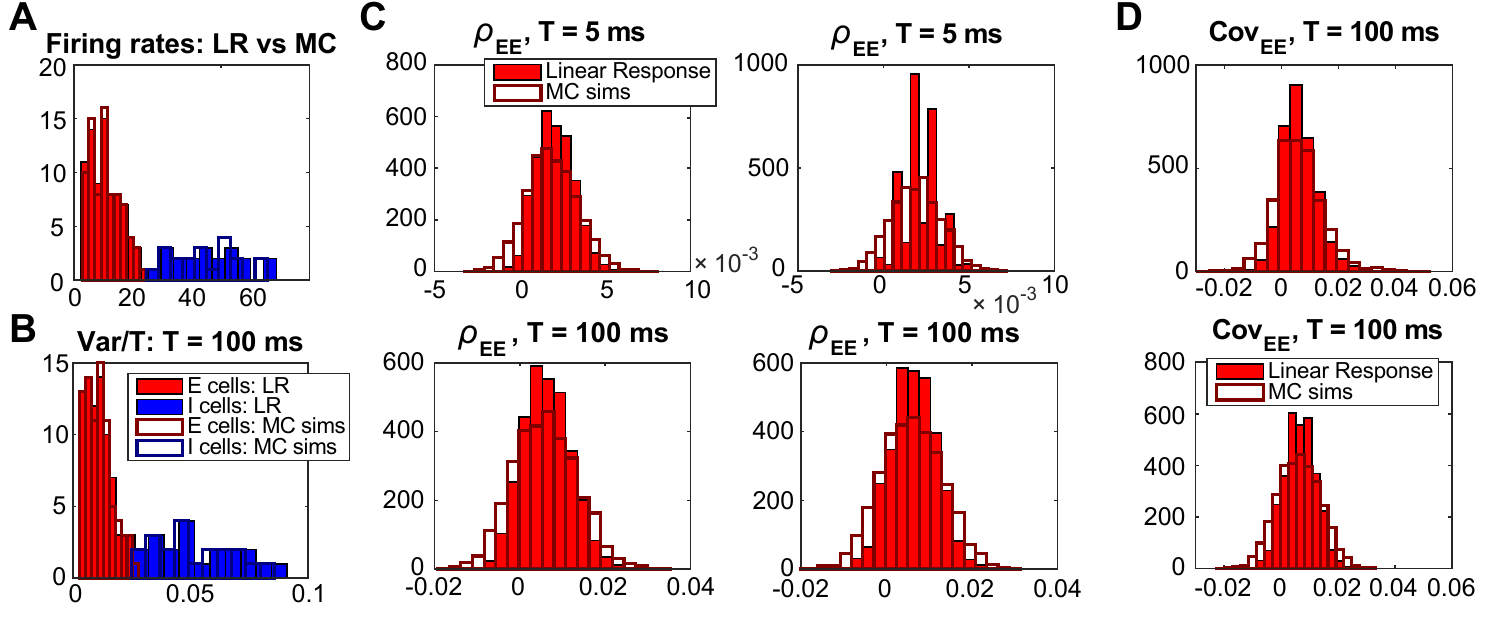}
 \caption{\label{fig:MC_vs_LR_stats_hist}  {\bf Theory predicts population statistics in the asynchronous regime.} Distributions of spiking statistics, comparing the results of linear response theory with Monte Carlo simulations in the asynchronous regime. (A) Firing rates $\nu_i$, for the heterogeneous network. (B) Normalized spike count variances,  $\Var_T[n_i]/T$, heterogeneous network, $T = $ 100 ms. (C) Spike count correlations, for 5 ms and 100 ms time windows: heterogeneous (left two panels) and homogeneous (right two panels). (D) Spike count \textit{covariance}, 100 ms time windows: heterogeneous (top) and homogeneous (bottom).} 
\end{figure}

We first consider  the firing rate, 
shown in Fig. \ref{fig:MC_vs_LR_stats_hist}A. In the heterogeneous network, both excitatory and inhibitory firing rates have large ranges that span approximately an order of magnitude.  
The linear response theory accurately captures all aspects of the firing rate distributions. The inhibitory firing rates are higher than excitatory, consistent with this population receiving a stronger excitatory input (vs. the excitatory population; compare $W_{IE}/N_E$ and $W_{EE}/N_E$, from Table 1 in the main text).
In the homogeneous network, firing rates are strongly clustered around their mean values (shown in Table \ref{table:statsAsyn}). They are also well-predicted, although the linear response theory does appear to slightly overestimate the inhibitory rates (see Table \ref{table:stats_MC_vs_LR}). Similarly, spike count variances match well (shown for long time windows ($T=100$ ms) in Fig. \ref{fig:MC_vs_LR_stats_hist}B).

\begin{table}[htb!]
\centering
\caption{
{\bf Statistics from heterogeneous vs. homogeneous networks: asynchronous regime}}
\begin{tabular}{ | l | c | c | c | c |  }
\hline
 &  \multicolumn{2}{|c|}{Heterogenous} & \multicolumn{2}{|c|}{Homogenous}\\
\hline
Statistic & E & I & E & I\\
\hline
Firing rate (Hz) & $10.6 \pm 5.0$ & $44.3 \pm 11.3$ &  $10.1 \pm 0.046$ & $43.5 \pm 0.37$\\
$\Var_T$, $T=5$ ms & $0.051 \pm 0.023$ & $0.19 \pm 0.048$& $0.048 \pm 0.0002$ & $0.19 \pm 0.0014$ \\
$\Var_T$, $T=100$ ms  & $1.14 \pm 0.58$ & $5.16 \pm 1.67$ & $1.06 \pm 0.0095$ &  $4.99 \pm 0.096$\\
\hline
& \multicolumn{2}{|c|}{Heterogenous} &  \multicolumn{2}{|c|}{Homogenous}\\
\hline
$\rho^{EE}$, $T=5$ ms & \multicolumn{2}{|c|}{$0.0019 \pm 0.0015$} & \multicolumn{2}{|c|}{$0.0019 \pm 0.0015$} \\
$\rho^{EE}$, $T=50$ ms & \multicolumn{2}{|c|}{$0.0058 \pm 0.0062$} & \multicolumn{2}{|c|}{$0.0060 \pm 0.0059$} \\
$\rho^{EE}$, $T=100$ ms & \multicolumn{2}{|c|}{$0.0059 \pm 0.0075$} & \multicolumn{2}{|c|}{$0.0059 \pm 0.0072$} \\ 
\hline
 \end{tabular}
\begin{flushleft}
Firing statistics from Monte Carlo simulations of recurrent networks in the asynchronous regime.
\end{flushleft}\label{table:statsAsyn}
\end{table} 

\begin{table}
\centering
\caption{
{\bf Statistics in recurrent networks: Monte Carlo vs. linear response theory, asynchronous regime}}
\begin{tabular}{ | l | c | c | c | c | c | c | c | c |  }
\hline
 &  \multicolumn{4}{|c|}{Heterogenous} & \multicolumn{4}{|c|}{Homogenous}\\
\hline
 &  \multicolumn{2}{|c|}{$\mu$} & \multicolumn{2}{|c|}{$\sigma$} &  \multicolumn{2}{|c|}{$\mu$} & \multicolumn{2}{|c|}{$\sigma$}\\
 \hline
 Statistic & MC & LR & MC & LR   & MC & LR & MC & LR \\
 \hline
Firing rate, E & $10.6$ & $10.6$ & $5.0$ & $5.3$ &$10.1$ & $10.0$ & $4.6 \times 10^{-2}$ & $2.3 \times 10^{-2}$ \\
Firing rate, I & $44.3$ & $45.9$ & $11.3$ & $12.0$ & $43.5$ &$45.0$& $0.37$ & $0.32$\\
FF, 5 ms, E & $0.9585$ & $0.9647$ & $0.0148$ & $0.0162$ &$0.9576$  & $0.9640$ & $4.53 \times 10^{-4}$ & $1.12 \times 10^{-4}$ \\
FF, 5 ms, I & $0.8725$ & $0.8726$ & $0.0093$ & $0.0091$ & $0.8690$ & $0.8688$ & $9.4 \times 10^{-4}$ & $5.81 \times 10^{-4}$ \\
FF, 100 ms, E & $1.0573$ & $1.0587$ & $0.0345$ & $0.0305$ & $1.0493$ & $1.0504$ & $0.0074$ & $0.0024$\\
FF, 100 ms, I & $1.1449$ & $1.1540$ & $0.0810$ & $0.0859$ & $1.1460$ & $1.1528$ & $0.0164$ & $0.0099$\\
\hline
$\rho^{EE}$, 5 ms ($\times 10^{-3}$)& $1.9$ & $2.0$& $1.5$ & $1.1$& $1.9$ & $2.1$ & $1.5$ & $1.0$\\
$\rho^{EE}$, 50 ms  ($\times 10^{-3}$) & $5.8 $ & $6.3$& $6.2 $ &$4.8$& $6.0 $ & $6.4$ & $5.9 $ & $4.6$\\
$\rho^{EE}$, 100 ms  ($\times 10^{-3}$) & $5.9$ & $6.3$& $7.5 $ &$5.3$& $5.9 $ & $6.4$& $7.2 $ & $5.3$\\
\hline
 \end{tabular}
\begin{flushleft}
Comparing Monte Carlo simulations with predictions from linear response; firing statistics in the asynchronous regime. Statistics displayed here are: firing rates for both excitatory and inhibitory populations; Fano factor (FF) for both excitatory and inhibitory populations; spike count correlations for excitatory pairs only ($\rho^{EE}$). Standard deviations are reported across the population; i.e. across eighty (80) E cells, or twenty (20) I cells, or 3160 E-E pairs.
\end{flushleft}\label{table:stats_MC_vs_LR} 
\end{table}

\begin{table}[htb!]
\centering
\caption{
{\bf Statistics from heterogeneous vs. homogeneous networks: strong asynchronous regime}}
\begin{tabular}{ | l | c | c | c | c |  }
\hline
 &  \multicolumn{2}{|c|}{Heterogenous} & \multicolumn{2}{|c|}{Homogenous}\\
\hline
Statistic & E & I & E & I\\
\hline
Firing rate (Hz) & $8.1 \pm 4.5$ & $36.6 \pm 9.8$ &  $7.2 \pm 0.095$ & $35.2 \pm 0.41$\\
$\Var_T$, $T=5$ ms & $0.039 \pm 0.021$ & $0.16 \pm 0.040$& $0.035 \pm 0.0004$ & $0.15 \pm 0.0016$ \\
$\Var_T$, $T=100$ ms  & $0.84 \pm 0.48$ & $3.93 \pm 1.31$ & $0.74 \pm 0.013$ &  $3.75 \pm 0.079$\\
\hline
& \multicolumn{2}{|c|}{Heterogenous} &  \multicolumn{2}{|c|}{Homogenous}\\
\hline
$\rho^{EE}$, $T=5$ ms & \multicolumn{2}{|c|}{$0.0119 \pm 0.0037$} & \multicolumn{2}{|c|}{$0.0109 \pm 0.0025$} \\
$\rho^{EE}$, $T=50$ ms & \multicolumn{2}{|c|}{$0.0622 \pm 0.0206$} & \multicolumn{2}{|c|}{$0.0587 \pm 0.0147$} \\
$\rho^{EE}$, $T=100$ ms & \multicolumn{2}{|c|}{$0.0654 \pm 0.0232$} & \multicolumn{2}{|c|}{$0.0618 \pm 0.0169$} \\ 
\hline
 \end{tabular}
\begin{flushleft}
Firing statistics from Monte Carlo simulations of recurrent networks in the strong asynchronous regime.
\end{flushleft}\label{table:statsSA}
\end{table} 

\begin{table}
\centering
\caption{
{\bf Statistics in recurrent networks: Monte Carlo vs. linear response theory, strong asynchronous regime}}
\begin{tabular}{ | l | c | c | c | c | c | c | c | c |  }
\hline
 &  \multicolumn{4}{|c|}{Heterogenous} & \multicolumn{4}{|c|}{Homogenous}\\
\hline
 &  \multicolumn{2}{|c|}{$\mu$} & \multicolumn{2}{|c|}{$\sigma$} &  \multicolumn{2}{|c|}{$\mu$} & \multicolumn{2}{|c|}{$\sigma$}\\
 \hline
 Statistic & MC & LR & MC & LR   & MC & LR & MC & LR \\
 \hline
Firing rate, E & $8.14$ & $6.84$ & $4.5$ & $4.2$ &$7.2$ & $6.0$ & $0.095$ & $0.057$ \\
Firing rate, I & $36.6$ & $36.4$ & $9.8$ & $9.9$ & $35.2$ &$34.9$& $0.41$ & $0.36$\\
FF, 5 ms, E & $0.9622$ & $0.9829$ & $0.0190$ & $0.0155$ &$0.9653$  & $0.9853$ & $4.80 \times 10^{-4}$ & $2.69 \times 10^{-4}$ \\
FF, 5 ms, I & $0.8719$ & $0.8802$ & $0.0147$ & $0.0136$ & $0.8709$ & $0.8788$ & $0.0014$ & $0.0012$ \\
FF, 100 ms, E & $1.0271$ & $1.0578$ & $0.0226$ & $0.0206$ & $1.0216$ & $1.0516$ & $0.0116$ & $0.0059$\\
FF, 100 ms, I & $1.0581$ & $1.0881$ & $0.0698$ & $0.0733$ & $1.0655$ & $1.0948$ & $0.0124$ & $0.0118$\\
\hline
$\rho^{EE}$, 5 ms ($\times 10^{-3}$)& $11.9$ & $8.0$& $3.7$ & $2.9$& $10.9$ & $7.6$ & $2.5$ & $1.9$\\
$\rho^{EE}$, 50 ms  ($\times 10^{-3}$) & $62.2 $ & $41.9$& $20.6 $ &$17.2$& $58.7 $ & $40.3$ & $14.7 $ & $12.0$\\
$\rho^{EE}$, 100 ms  ($\times 10^{-3}$) & $65.4$ & $44.2$& $23.2$ &$19.4$& $61.8 $ & $42.8$& $16.9 $ & $14.0$\\
\hline
 \end{tabular}
\begin{flushleft}
Comparing Monte Carlo simulations with predictions from linear response; firing statistics in the strong asynchronous regime. Statistics displayed here are: firing rates for both excitatory and inhibitory populations; Fano factor (FF) for both excitatory and inhibitory populations; spike count correlations for excitatory pairs only ($\rho^{EE}$). Standard deviations are reported across the population; i.e. across eighty (80) E cells, or twenty (20) I cells, or 3160 E-E pairs.
\end{flushleft}\label{table:stats_MC_vs_LR_SA} 
\end{table}

We now consider a common measure of noise correlations, the spike count (Pearson's) correlation of pairs of excitatory cells in a particular time window (Eqn. (40), main text).
As in the Monte Carlo simulations, we have assumed spike count statistics to be stationary over time, so that for each $T$, spike counts $n_i$ and $n_j$ are treated as random variables sampled both over realizations (i.e trials) and time $t$.  
We computed these statistics for both short ($T=5$ ms) and long ($T=100$ ms) time windows, 
and illustrate them in Figure \ref{fig:MC_vs_LR_stats_hist}C; statistics from both heterogeneous (left panels) and homogenous (right panels) are show. E-E correlations are weakly positive, with a small fraction 
of pairs ($\sim$ 5\%) having values below zero. 
In all panels, the mean/median of the distribution are captured well by the linear response theory; however, the linear response calculation appears to slightly underestimate the simulated variance, as evidenced by the ``taller and thinner" distribution shown in solid red (each histogram is computed by distributing $80 \times 79/2$ distinct coefficients over equally sized bins). 
The correlation values computed by linear response have comparable 
ranges in the heterogeneous and homogenous networks, similar to MC simulations and in contrast to first-order statistics.

Linear response theory also predicts the distribution of spike count \textit{covariances} (i.e. the numerator of Eqn. (40)): we show these in Fig. \ref{fig:MC_vs_LR_stats_hist}D. As for correlations, theory underestimates the observed variance of the distributions. However, it appears to capture the fat right tails in the heterogeneous network very well (Fig. \ref{fig:MC_vs_LR_stats_hist}D, top row).

\begin{figure}
\centering
\includegraphics[width=\textwidth]{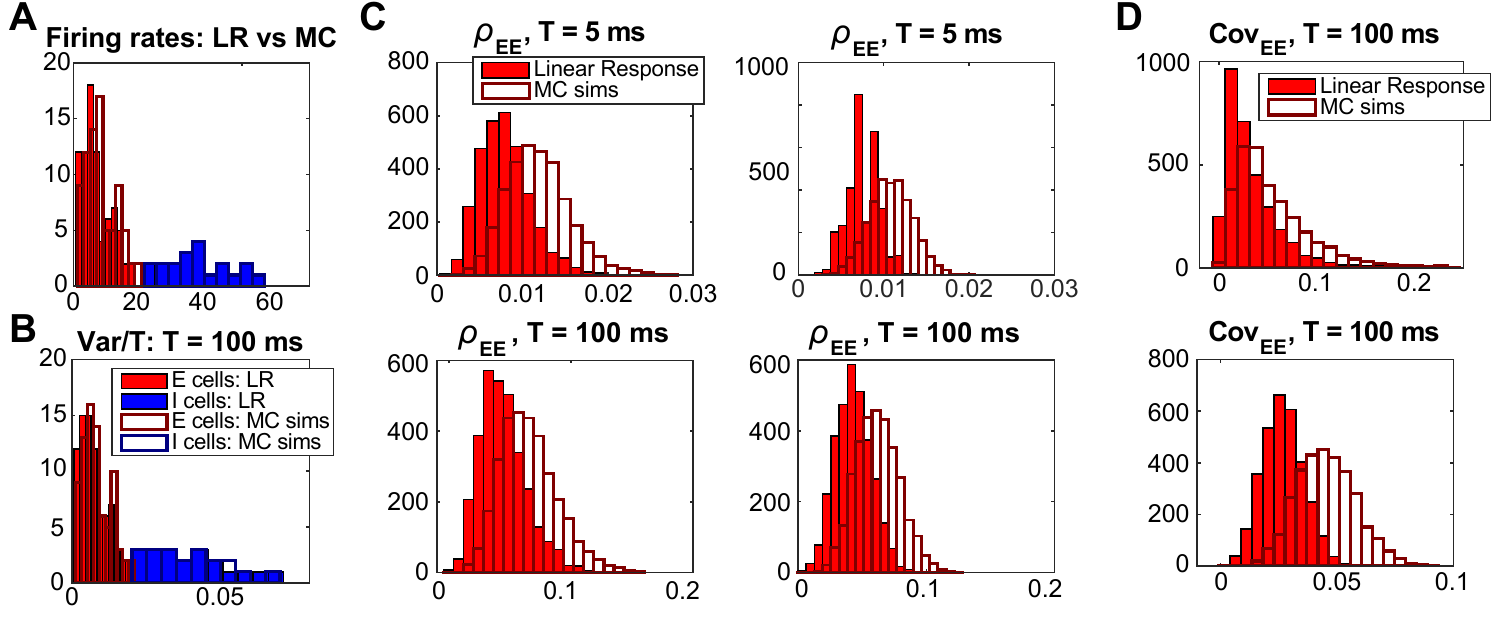}
 \caption{\label{fig:MC_vs_LR_stats_hist_Str}   {\bf Theory predicts population statistics in the strong asynchronous regime.} Distributions of spiking statistics, comparing the results of linear response theory with Monte Carlo simulations in the \textit{strong asynchronous} regime. (A) Firing rates $\nu_i$, for the heterogeneous network. (B) Normalized spike count variances,  $\Var_T[n_i]/T$, heterogeneous network, $T = $ 100 ms. (C) Spike count correlations, for 5 ms and 100 ms time windows: heterogeneous (left two panels) and homogeneous (right two panels). (D) Spike count \textit{covariance}, 100 ms time windows: heterogeneous (top) and homogeneous (bottom).} 
\end{figure}

We now turn our attention to the \textit{strong asynchronous} ({\bf SA}) regime, in which both types of excitatory connections were strengthened (see Table 1);
the resulting network shows occasional, irregular bursts of concentrated activity (see Fig. 1B).
Many of the overall trends are similar to the asynchronous case; we focus on the differences.

Excitatory firing rates were slightly under-predicted by linear response theory (Fig. \ref{fig:MC_vs_LR_stats_hist_Str}A; see Table \ref{table:statsSA} for homogeneous rates).
Similarly, spike count variances (Fig. \ref{fig:MC_vs_LR_stats_hist_Str}B) were under predicted.
Spike count correlations $\rho^{EE}$ are now positive, with few or no negative correlations  (Fig. \ref{fig:MC_vs_LR_stats_hist_Str}C). The mean is significantly under-predicted; the predicted distributions appear slightly narrower than the observed (Monte Carlo) distribution.
Spike count covariances for long time windows are shown in Fig. \ref{fig:MC_vs_LR_stats_hist_Str}D; the linear response theory appears to capture the qualitative shape of the distributions, particularly the fat right tail in the heterogeneous network (top panel). However, as for correlations (Fig. \ref{fig:MC_vs_LR_stats_hist_Str}C), the mean is under-predicted in both networks (Table \ref{table:stats_MC_vs_LR_SA}).

\label{S3_Appendix}

\subsection*{Linear response theory predicts the first- and second-order statistics of individual cells}

We next investigate how well these statistics are predicted on a \textit{cell-to-cell} basis. This is crucially important when individual correlation coefficients $\rho_{ij}$ within a simulation may vary over an order of magnitude or even in sign.  For example, consider the heterogeneous network illustrated in Fig. 2C(bottom): 
E-E correlations were weakly positive (on average less than 0.01) but could range as high as 0.03 or as negative as -0.015 for some cell pairs. If I pick a specific cell pair $i$, $j$ out of the population, can I predict where in this range $\rho_{ij}$ will fall? Predicting the correlation of specific cell pairs would be a valuable tool, as many models for heterogeneity --- such as those based on population density methods (e.g. \cite{Ly_intrsNet_15} ) --- do not capture cell-to-cell variation.

\begin{figure}
\centering
\includegraphics[width=\textwidth]{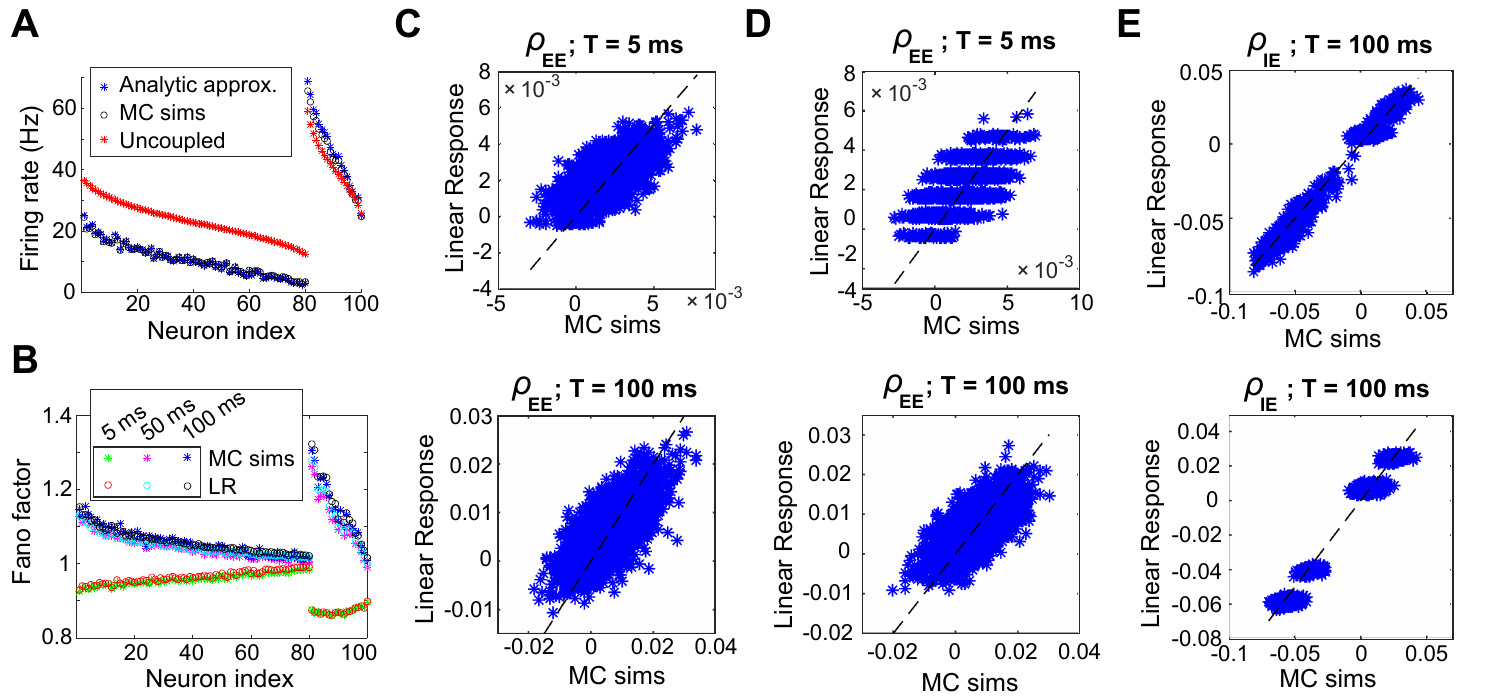}
\caption{\label{fig:MC_vs_LR_asyn_single_cell_stats} 
 {\bf Theory predicts cell-by-cell statistics in the asynchronous regime.}
Distributions of spiking statistics, comparing the results of linear response theory with Monte Carlo simulations in the asynchronous regime, cell-by-cell. (A) Firing rates $\nu_i$, for the heterogeneous network. (B)  Fano factor, heterogeneous network. Time windows ($T$) shown are: $5$ ms, $50$ ms, $100$ ms. (C) Spike count correlations, heterogeneous network. Time windows are: 5 ms (top) and 100 ms (bottom).
(D) Spike count correlations, homogenous network. Time windows are: 5 ms (top) and 100 ms (bottom).
(E) E-I spike count correlations, $T=100$: heterogenous (top) and homogeneous (bottom). } 
\end{figure}

We find that single-cell statistics are very accurately predicted for the heterogeneous network. In Figure \ref{fig:MC_vs_LR_asyn_single_cell_stats}, we show firing rate (Fig. \ref{fig:MC_vs_LR_asyn_single_cell_stats}A) 
and Fano factor (Fig. \ref{fig:MC_vs_LR_asyn_single_cell_stats}B) for three time windows: $T = 5, 50$, $100$ ms.  In each panel, both quantities from the Monte Carlo simulations and linear response theory are plotted, on a cell-by-cell basis. In Fig.  \ref{fig:MC_vs_LR_asyn_single_cell_stats}A, the red stars give the firing rate of the uncoupled neurons (i.e. determined only by the threshold $\theta_i$ and the level of background noise).  The effect of coupling is to lower the firing rate of the E cells but to raise the firing rate of the I cells; this is captured very well by the fixed point iteration of Eqn.  (22).
There is still significant heterogeneity in the firing rates due to variable threshold, with high threshold neurons maintaining comparatively lower firing rates and low threshold neurons maintaining comparatively higher firing rates.

We now analyze the ability of linear response to predict \textit{two-cell} statistics. In Fig.  \ref{fig:MC_vs_LR_asyn_single_cell_stats}C
we plot the 
spike count correlation $\rho_{ij} = \Cov_T(n_i,n_j)/\sqrt{\Var[n_i] \Var[n_j]}$, for all possible E-E pairs in the heterogeneous network, at both $T= 5$ ms (top) and $100$ ms (bottom).
The values predicted by linear response theory matches well with the Monte Carlo simulations in both overall range and cell-to-cell; in both plots, the points cluster around the unity line.

We now consider how well linear response models the homogeneous network on a cell-to-cell basis. 
As in the heterogeneous network, single-cell statistics are accurately predicted (because both simulated and predicted single-cell statistics are nearly constant across the population, we report their values in Table \ref{table:stats_MC_vs_LR}).
Firing rate is slightly overestimated, as is variance. 
Fano factor differs systemically with time interval: cell activities appear slightly ``sub-Poisson" for $T = 5$ ms, but ``super-Poisson" 
for $T= 50, 100$ ms. We then examined two-cell statistics: E-E correlations were weak and positive, and clustered in a cloud around the unity line (Fig.  \ref{fig:MC_vs_LR_asyn_single_cell_stats}D), for both short ($T=$5 ms, top) and long ($T=$100 ms, bottom) time windows.

\begin{figure}
\centering
\includegraphics[width=\textwidth]{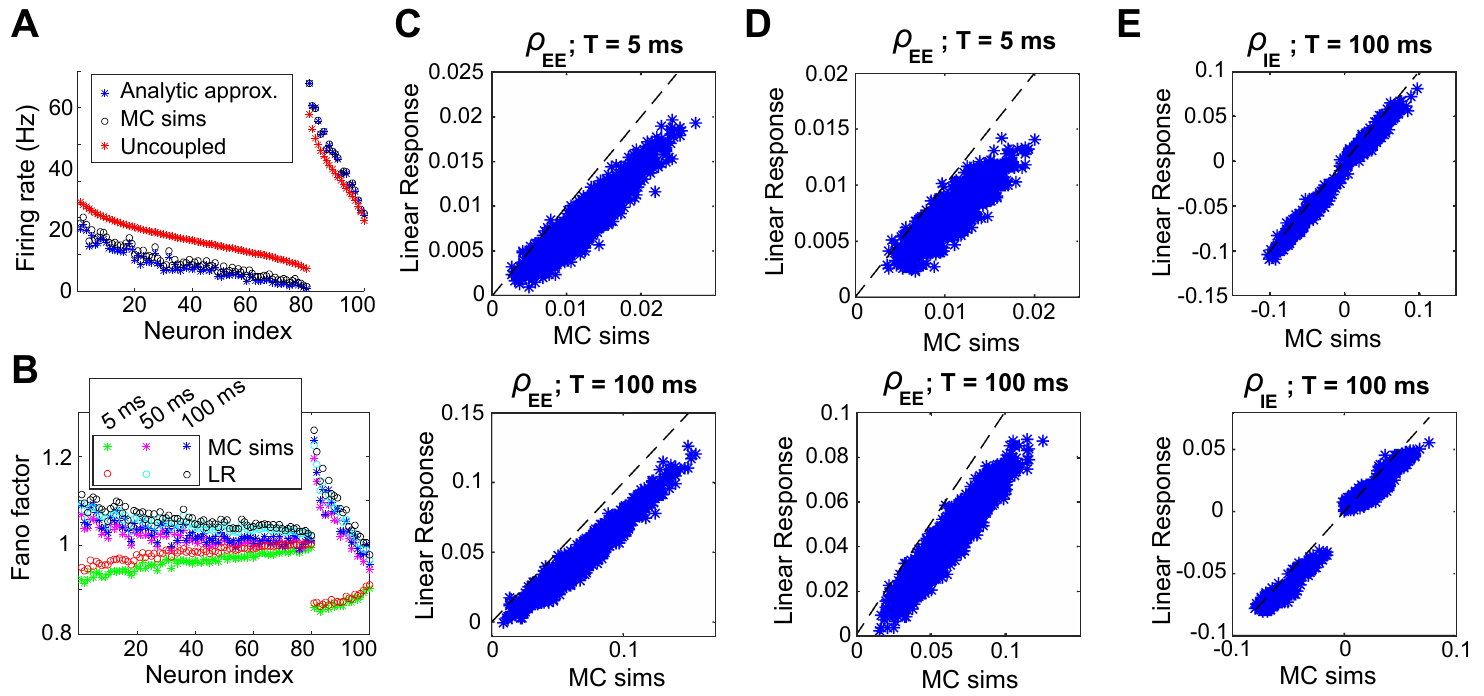}
\caption{\label{fig:MC_vs_LR_Str_single_cell_stats} 
 {\bf Theory predicts cell-by-cell statistics in the strong asynchronous regime.}
Comparing the results of linear response theory with Monte Carlo simulations in the \textit{strong asynchronous} regime, cell-by-cell. 
(A) Firing rates $\nu_i$, for the heterogeneous network. (B)  Fano factor, heterogeneous network. Time windows ($T$) shown are: $5$ ms, $50$ ms, $100$ ms. (C) Spike count correlations, heterogeneous network. Time windows are: 5 ms (top) and 100 ms (bottom).
(D) Spike count correlations, homogenous network. Time windows are: $5$ ms (top) and $100$ ms (bottom).
(E) E-I spike count correlations, $T=100$: heterogenous (top) and homogeneous (bottom).} 
\end{figure}

Although we mostly focus on E-E correlations here, we observed excellent results in predicting other two-cell statistics, for example excitatory-inhibitory (E-I) correlations.  In Fig.  \ref{fig:MC_vs_LR_asyn_single_cell_stats}E we show E-I correlations for $T=100$ ms, for both the heterogeneous (top) and homogeneous (bottom) networks. E-I correlations took on a wider range of values; both positive and negative, with a range between $[-0.15, 0.15]$ for $T=100$ ms. In the homogeneous network they cluster in four distinct clouds (Fig.  \ref{fig:MC_vs_LR_asyn_single_cell_stats}E, bottom): on closer inspection, these correspond to the presence or absence of direct connections between the pairs. For E-I pairs with no direct connection, correlations are weak and positive. Pairs with only a $E \rightarrow I$ connection are strongly positively correlated, while pairs with only an $I \rightarrow E$ connection are strongly negatively correlated. Pairs with BOTH connections are weakly negatively correlated, which may reflect the fact that $W_{IE} > W_{EI}$. 
We also find good results when we move to the strongly asynchronous case. This network has increased excitation ($W_{EE} = 9$ and $W_{EI}= 8$, vs. $W_{EE} = 5$ and $W_{EI}= 0.5$ in the asynchronous regime) and shows short bursts of activity 
(see Fig. 1);
since this violates the assumption of constant firing rate, \textit{a priori} we cannot be sure linear response theory will be successful.  However, the theory is nonetheless successful at matching broad trends in firing rate, Fano factors, and cell-pair correlations (Fig. \ref{fig:MC_vs_LR_Str_single_cell_stats}). 
There are differences between the simulations and linear response calculations. 
For excitatory neurons, firing rate is slightly overestimated (Fig. \ref{fig:MC_vs_LR_Str_single_cell_stats}A), variance underestimated and Fano factor overestimated  (Fig. \ref{fig:MC_vs_LR_Str_single_cell_stats}B). For inhibitory neurons, firing rate appears to be very accurate; variance and Fano factor are slightly overestimated.
We also see that $\rho^{EE}$ is systematically underestimated (Fig. \ref{fig:MC_vs_LR_Str_single_cell_stats}C, heterogeneous;  Fig. \ref{fig:MC_vs_LR_Str_single_cell_stats}D, homogeneous;); $\rho^{IE}$ may also be slightly underestimated, but less so (see Fig. \ref{fig:MC_vs_LR_Str_single_cell_stats}E).

\begin{figure}
\centering
  \includegraphics[width=\textwidth]{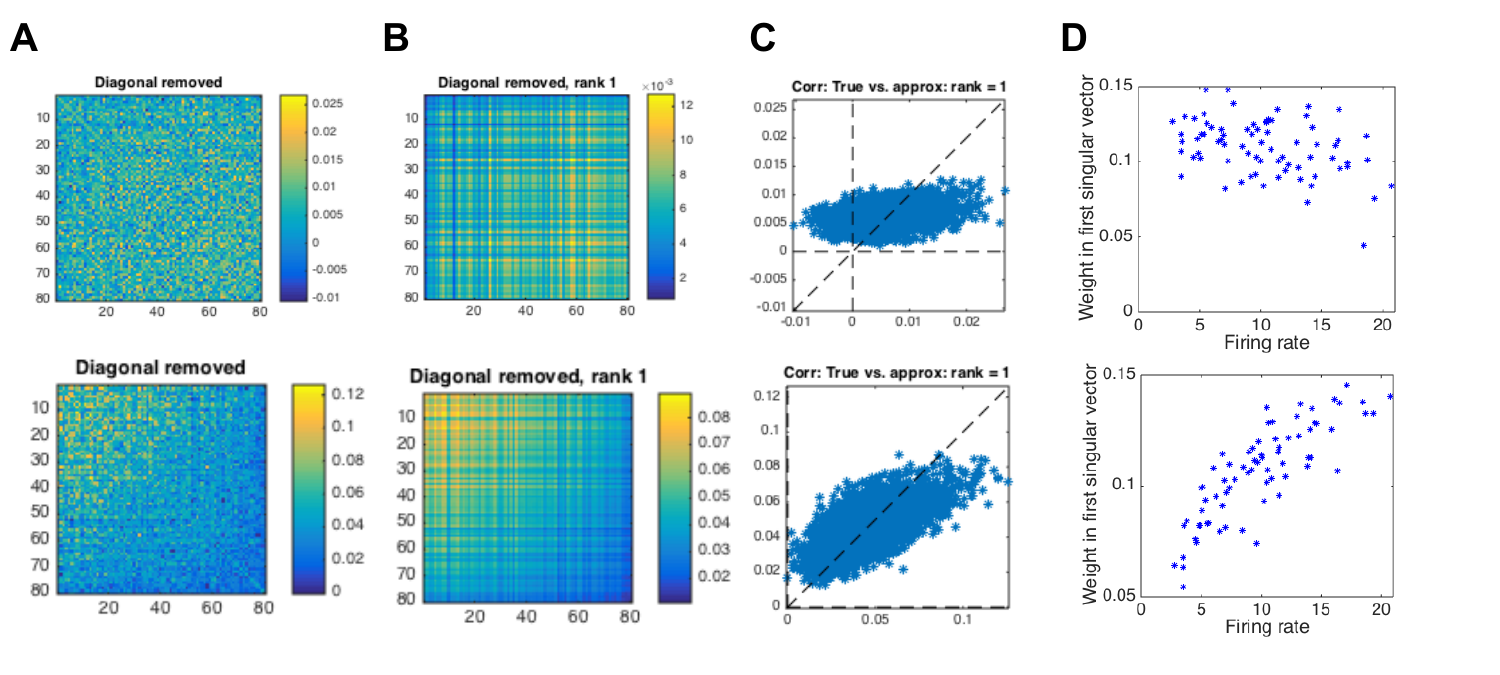}
\caption{\label{fig:lowRank_Asyn_vs_Str_Het_LR} {\bf Theory captures low-rank structure in correlation matrices.} Approximating correlation matrices (for the heterogeneous networks) obtained from linear response theory, as a diagonal plus rank-one.  In each column of (A-D), the asynchronous (top) and strong asynchronous (bottom) regimes are shown; $T = $ 100 ms.  (A) The shifted E-E correlation matrix, $\Cvec_T-\lambda \Ivec$, for an appropriately chosen $\lambda$. (B) A rank-one approximation to $\Cvec_T-\lambda \Ivec$. (C) True correlation coefficients vs. rank-one approximation, cell-by-cell. (D) Weight in the first singular vector, $\uvec_1$ vs.geometric mean firing rate $\sqrt{\nu_i \nu_j}$.} 
\end{figure}
Finally, the cell-by-cell accuracy of the linear response theory is reflected in the overall structure of the correlation matrix. We performed the diagonal plus rank-one analysis on correlation matrices we obtained from linear response theory (Fig. \ref{fig:lowRank_Asyn_vs_Str_Het_LR}). 
We see the same patterns observed in Fig 4; 
in the strong asynchronous regime  there is a strong positive relationship with firing rate, which is reflected in the weights of the first singular vector.  

\clearpage

\label{S4_Appendix}
\subsection*{Approximating single-cell susceptibility in a heterogeneous network}

To understand how the single-cell susceptibility (Eqn. (4), main text) depends on the six parameters $\langle g_{I,i} \rangle$, etc., we plotted each parameter vs. firing rate (see Fig. S6 and Fig. S7 for the asynchronous and strong asynchronous regimes respectively). The panels in Fig. S6A show $\langle g_{I,i} \rangle$, $\sigma_{I,i}$, $\langle g_{E,i} \rangle$, and $\sigma_{E,i}$, which appear to be randomly scattered with no relationship to firing rate (there is also no apparent relationship in the derived relationships for effective time constant $\tau_{{\rm eff},i}$, effective potential $\mu_i$, and effective noise $\sigma_{{\rm eff},i}$, Fig. S6B);  $\sigma_i$ is constant for all cells.
However, for both networks there is a clear relationship with $\theta_i$.  Furthermore, of the four non-constant parameters with no discernible relationship, the values  of $\langle g_I \rangle$ appear to have the greatest spread; we therefore hypothesized that we can approximate $S^{\langle g_I \rangle}_{i}$, by reevaluating the firing rate function in which $\sigma_{I,i}$, $\langle g_{E,i} \rangle$, $\sigma_{E,i}$ and  $\sigma_i$ have been replaced by their average values: i.e. 
\begin{eqnarray}
\hat{S}^{\langle g_I \rangle}_{i} & \equiv &\frac{1}{\sqrt{F( \langle g_{I,i} \rangle, \theta_i)} }  \frac{\partial F}{\partial x_1}\left( \langle g_{I,i} \rangle, \theta_i \right)   \label{eqn:susc_fix_param}
\end{eqnarray}
 where
 \begin{eqnarray}
F( \langle g_{I,i} \rangle, \theta_i) & \equiv &  f \left( \langle g_{I,i} \rangle, \langle \sigma_{I,i} \rangle_p, \langle \, \langle g_{E,i} \rangle \, \rangle_p, \langle \sigma_{E,i} \rangle_p, \langle \sigma_i \rangle_p, \theta_i \right)   \label{eqn:F_def}
\end{eqnarray}
and  $\langle \, \cdot \, \rangle_p$ denotes the population average.  

To explore the role of how the \textit{cause} of firing rate diversity might regulate correlations, we next plot firing rate as a function of threshold $\theta$ and  mean inhibitory conductance $\langle g_I \rangle$; that is, we plot $F( \langle g_{I} \rangle, \theta)$ (see Eqn.  \eqref{eqn:F_def}). In both regimes, firing rate decreases from left to right and bottom to top (Fig. \ref{fig:diversity_mech_rate_2D}). Any curve transversal to the level curves of the firing rate will sample a wide range of firing rates, but possibly different susceptibilities. The points corresponding to the actual excitatory cells in our network are illustrated in red; black squares illustrate an alternate curve, where  $\langle g_I \rangle$ is varied but $\theta = 1$.

\begin{figure}
\centering
 \includegraphics[width=\textwidth]{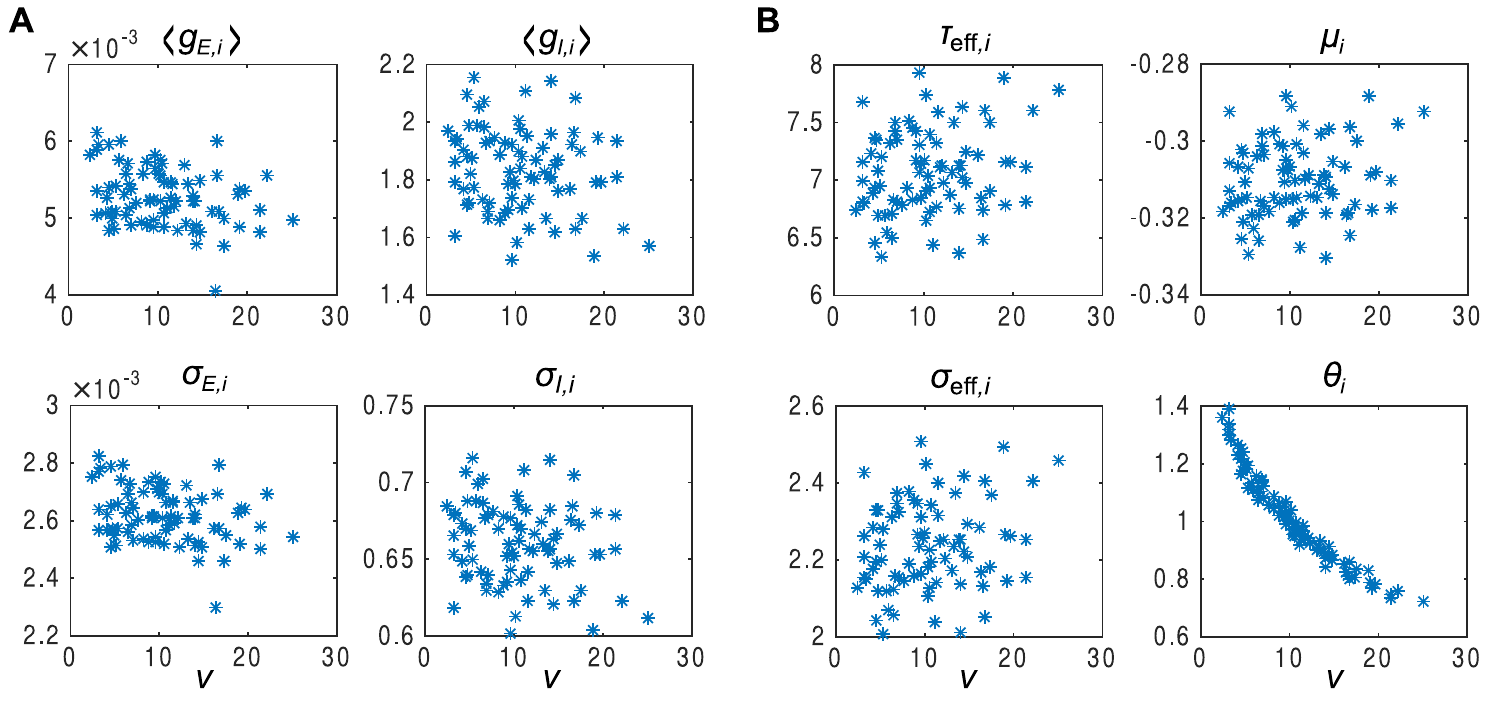}
\caption{{\bf Effective parameters in the heterogeneous network: asynchronous regime} Parameters used to estimate susceptibility, for all excitatory neurons in the network. Each parameter is plotted vs. firing rate. (A) Mean excitatory conductance $\langle g_{E,i} \rangle$ (top left), mean inhibitory conductance $\langle g_{I,i} \rangle$ (top right), excitatory conductance variability $\sigma_{E,i}$ (bottom left), and inhibitory conductance variability $\sigma_{I,i}$ (bottom right). (B) Effective time constant $\tau_{{\rm eff},i}$ (top left), effective input current $\mu_i$ (top right), effective current noise variability $\sigma_{{\rm eff},i}$ (bottom left), and threshold $\theta_i$ (bottom right). }
\end{figure}

\begin{figure}
\centering
 \includegraphics[width=\textwidth]{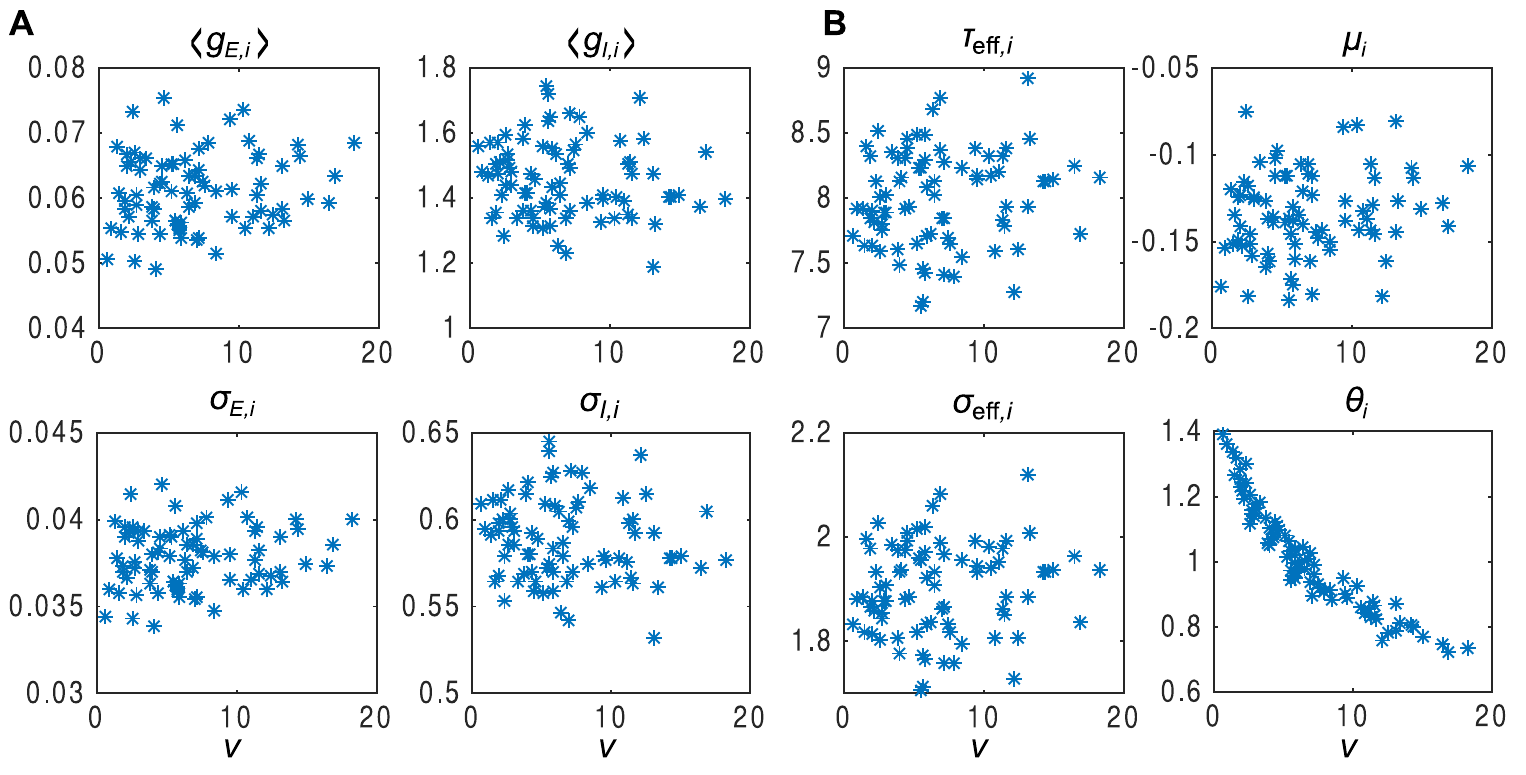}
\caption{{\bf Effective parameters in the heterogeneous network: strong asynchronous regime} Parameters used to estimate susceptibility, for all excitatory neurons in the network. Each parameter is plotted vs. firing rate. (A) Mean excitatory conductance $\langle g_{E,i} \rangle$ (top left), mean inhibitory conductance $\langle g_{I,i} \rangle$ (top right), excitatory conductance variability $\sigma_{E,i}$ (bottom left), and inhibitory conductance variability $\sigma_{I,i}$ (bottom right). (B) Effective time constant $\tau_{{\rm eff},i}$ (top left), effective input current $\mu_i$ (top right), effective current noise variability $\sigma_{{\rm eff},i}$ (bottom left), and threshold $\theta_i$ (bottom right). }
\end{figure}

\begin{figure}
\centering
\includegraphics[width=\textwidth]{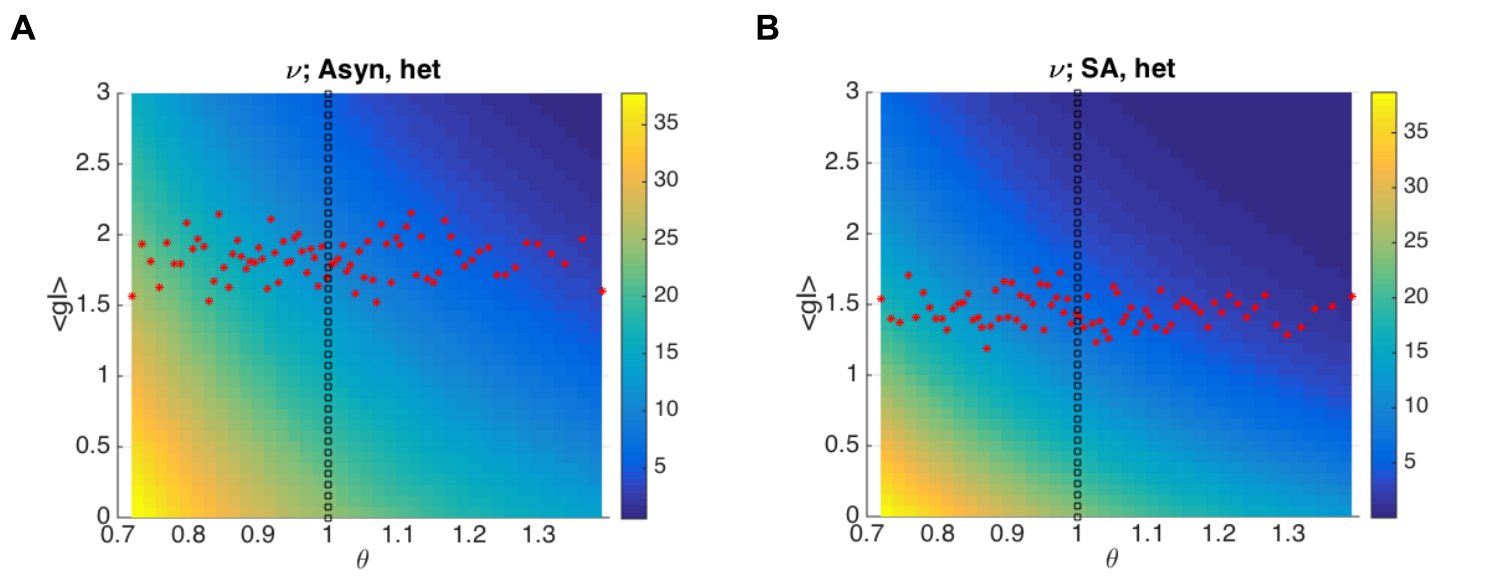}
\caption{{\bf Firing rate as a function of inhibitory conductance and threshold.} Firing rate of a conductance-based LIF neuron, as a function of mean inhibitory conductance $\langle g_I \rangle$ and threshold $\theta$: $\hat{S}^{\langle g_I \rangle} (\langle g_I \rangle, \theta)$ (defined in Eqn (19)). Other parameters are set to the population average.  Overlays show $(\langle g_{I,i} \rangle, \theta_i)$ values of the actual cells in the network (red stars) and an alternative curve through the plane, $(\langle g_{I} \rangle, 1)$, along which comparable firing rate diversity can be observed (black squares). (A) Asynchronous regime. (B) Strong asynchronous regime.} \label{fig:diversity_mech_rate_2D}
\end{figure}

\end{document}